\documentclass[a4paper,fleqn,usenatbib]{mnras}

\usepackage{mathptmx}
\usepackage[T1]{fontenc}
\usepackage{ae,aecompl}
\usepackage{amsmath}
\usepackage{amssymb}
\usepackage{graphicx}
\usepackage{ulem}

\usepackage[utf8]{inputenc}
\usepackage{longtable}
\usepackage{lscape}
\usepackage{marvosym}

\newcommand{\noc}[1]{{#1}}				

\title[helicity inversion mechanism]{Helicity inversion in spherical convection as a means for equatorward dynamo wave propagation}

\author[Duarte et al.]{
Lúcia D. V. Duarte,$^{1}$\thanks{E-mail: lduarte@astro.ex.ac.uk}
Johannes Wicht,$^{2}$
Matthew K. Browning,$^{1}$
\newauthor
and 
Thomas Gastine$^{2}$
\\
$^{1}$Department of Physics and Astronomy, University of Exeter, UK\\
$^{2}$Max-Planck-Institut für Sonnensystemforschung, G\"ottingen, Germany
}

\date{Accepted XXX. Received YYY; in original form ZZZ}

\pubyear{2015}

\begin{document}
\label{firstpage}
\pagerange{\pageref{firstpage}--\pageref{lastpage}}
\maketitle

\begin{abstract}

We discuss here a purely hydrodynamical mechanism to invert the sign of the kinetic helicity, which plays a key role in determining the direction of propagation of cyclical magnetism in most models of dynamo action by rotating convection. Such propagation provides a prominent, and puzzling constraint on dynamo models. In the Sun, active regions emerge first at mid-latitudes, then appear nearer the equator over the course of a cycle, but most previous global-scale dynamo simulations have exhibited poleward propagation (if they were cyclical at all). Here, we highlight some simulations in which the direction of propagation of dynamo waves is altered primarily by an inversion of the kinetic helicity throughout much of the interior, rather than by changes in the differential rotation. This tends to occur in cases with a low Prandtl number and internal heating, in regions where the local density gradient is relatively small. We analyse how this inversion arises, and contrast it to the case of convection that is either highly columnar (i.e., rapidly rotating) or locally very stratified; in both of those situations, the typical profile of kinetic helicity (negative throughout most of the northern hemisphere) instead prevails.

\end{abstract}

\begin{keywords}
convection -- dynamo -- hydrodynamics -- magnetic fields -- turbulence -- Sun: general -- stars: general
\end{keywords}

\section{Introduction}
\label{intro}

The systematic equatorward migration of sunspot emergence latitudes constitutes one of the most enduring puzzles of solar physics. Spots appear first at mid-latitudes, then progressively nearer the equator over the course of a roughly 11-year cycle (e.g., \citeauthor{Carrington58} \citeyear{Carrington58}; \citeauthor{Maunder04} \citeyear{Maunder04}; reviews in \citeauthor{Ossendrijver03} \citeyear{Ossendrijver03}, \citeauthor{Hathaway10a} \citeyear{Hathaway10a}), constituting the famous ``butterfly diagram". There is now widespread agreement that the surface magnetism ultimately arises from the action of a dynamo within the Sun's electrically conducting convection zone, but a detailed explanation for the equatorward propagation has remained elusive \citep[see, e.g.,][]{Moffatt78}. In many models of the global solar dynamo, this propagation is taken to reflect underlying wave-like behaviour in the generation of sub-surface magnetism \citep[see, e.g., review in][]{Charbonneau10,Priest14}.

In the classic model developed by \cite{Parker55} and explored by many other authors since \citep[e.g.,][]{Yoshimura75,Stix76,Gilman83}, the wave-like behaviour arises from the combination of helical turbulence and differential rotation, described in mean-field theory by the $\alpha$-effect and $\Omega$-effect, respectively (e.g., \citeauthor{Steenbeck66} \citeyear{Steenbeck66}; review in \citeauthor{Brandenburg05} \citeyear{Brandenburg05}). The former encapsulates the production of poloidal magnetic field from toroidal (or vice versa) by convective eddies that rise and twist, while the latter describes the generation of toroidal field by linear winding of a poloidal field by differential rotation. In an $\alpha\Omega$ dynamo, the sign of newly generated toroidal field is determined by the sense of the differential rotation and by the sign of the pre-existing poloidal field. The latter in turn depends on the properties of the convective flows. The direction of propagation of the dynamo wave is then determined by the locations where toroidal field, generated by the differential rotation from stretching of poloidal field loops, cancels or enhances the pre-existing toroidal field: if the newly generated field tends preferentially to cancel pre-existing field near the equator and reinforce it at higher latitudes, the migration of the field will be poleward. This is encapsulated by the well-known Parker-Yoshimura sign rule that dynamo waves in such models travel in a direction given by $\mathbf{s} = \alpha\nabla\Omega\times\vec{e_\phi}$ \citep{Yoshimura75,Stix76}.

The sign of the $\alpha$-effect, which partly determines the direction of propagation of the dynamo wave, is fundamentally related to the lack of reflectional symmetry in the flow: $\alpha$ changes sign under transformations from right-handed to left-handed coordinate systems, and vanishes if the velocity field is statistically invariant under such parity transformations \citep[see, e.g.,][]{Moffatt78}. In many variants of mean field theory, assuming certain simplifying features about the velocity field, $\alpha$ can in turn be related to the kinetic helicity of the flow, $\mathbf{u}\cdot\mathbf{\omega} = \mathbf{u}\cdot(\nabla\times\mathbf{u})$ (with $\mathbf{\omega}$ the vorticity), often in addition to other terms involving the current helicity \citep[e.g.,][]{Pouquet76,Brandenburg05}. The sign and spatial variation of the kinetic helicity are thus crucial for determining the direction of field propagation.

In the past few decades, a wide variety of published nonlinear dynamo simulations in spherical geometries have been shown to exhibit cyclical behaviour (e.g., \citeauthor{Gilman83} \citeyear{Gilman83}; \citeauthor{Ghizaru10} \citeyear{Ghizaru10}, in the context of stellar astrophysics, or \citeauthor{Goudard08} \citeyear{Goudard08}; \citeauthor{Schrinner11} \citeyear{Schrinner11}; \citeauthor{Simitev12} \citeyear{Simitev12}; \citeauthor{Gastine12a}; \citeyear{Gastine12a} in the planetary context). However, to the extent that these simulations have exhibited systematic latitudinal propagation, this has generally been poleward \citep[see, e.g., discussion in][]{Brun13}, in agreement with the Parker-Yoshimura rule (given the realized kinetic helicity and differential rotation) but in conflict with the observed solar butterfly diagram. More recently, a few groups have published examples of convective dynamos whose propagation is equatorward: e.g., \cite{Kapyla12,Kapyla13,Augustson13,Warnecke14}. In each of these cases, the equatorward migration has been attributed largely to features in the differential rotation: e.g., to regions where the radial $\Omega$ gradient is negative \citep[][]{Kapyla12,Warnecke14}, or to non-linear feedbacks on the shear \citep[][]{Augustson15}. The kinetic helicity profile in these simulations, and with it the purported $\alpha$-effect, appears to be largely as described in Sec.~\ref{helicity} below, and as realized in many previous simulations: it is negative in the northern hemisphere, leading to a positive $\alpha$-effect.

In this paper, we explore the circumstances under which the kinetic helicity, and with it the generation of poloidal magnetic fields, actually has this spatial distribution. In Sec.~\ref{scenarios} we review the processes that lead to this helicity distribution, and outline a scenario in which it could instead have the opposite sign throughout much of the spherical domain. In Sec.~\ref{mod}, we carry out non-linear simulations of anelastic convective dynamos in global spherical shells, and show that these can indeed exhibit such ``reversed" kinetic helicity profiles in certain regimes. We demonstrate that simulations exhibiting this reversal also show systematic equatorward propagation of dynamo waves, without any accompanying changes in the differential rotation. We analyse the mechanisms behind the kinetic helicity reversal more thoroughly in Sec.~\ref{results1}, and close in Sec.~\ref{results2} with a summary of our work and a discussion of its possible relevance to the Sun, other stars, and planets.

\section{Regimes of kinetic helicity}
\label{scenarios}

\subsection{Classical helicity configuration in global dynamo simulations}
\label{helicity}

The kinetic helicity $H$, calculated as
\begin{equation}
	H = \mathbf{u}\cdot\mathbf{\omega} = \mathbf{u}\cdot(\nabla\times\mathbf{u}) \textrm{,}
\label{eq:kinhel}
\end{equation}
arises from the twisting and writhing of convective flows as they rise or fall. It is instructive to consider two regimes of convection that have been the subject of particularly wide study, which we will call ``columnar" and ``plume-like". The former has been widely studied in the planetary context \citep[see, for example][]{Busse70,Busse77,Olson99,Busse02,Aubert05,Aubert08} and found to dominate in rapidly-rotating systems at lower Rossby numbers. In Boussinesq models, the transition to such behaviour has been discussed by \cite{Soderlund12} and we will address this matter in Sec.~\ref{column}.

\begin{figure}
\begin{center}
{\centering
	\includegraphics[trim=0cm 0cm 0.5cm 0cm, height=2.3in]{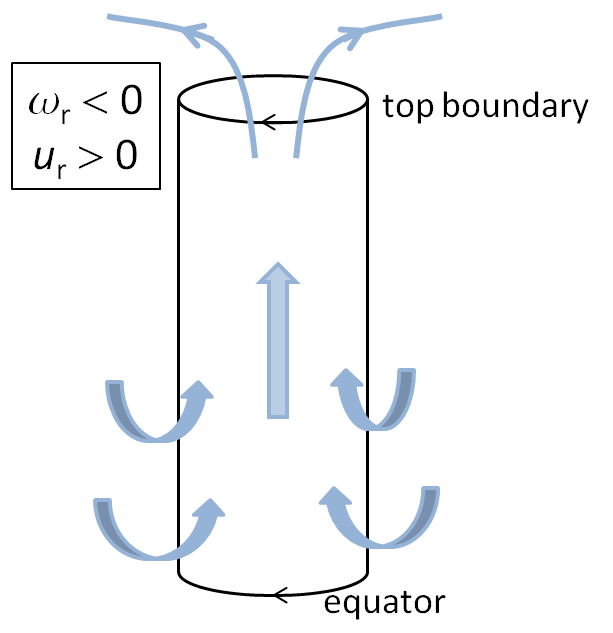}
	}
\caption{\small Diagram of vorticity generation in columnar convection, in a rotating spherical shell.
\label{fig:column2}}
\end{center}
\end{figure}

\begin{figure}
\begin{center}
{\centering
	\includegraphics[trim=0cm 0cm 0.5cm 0cm, height=2.9in]{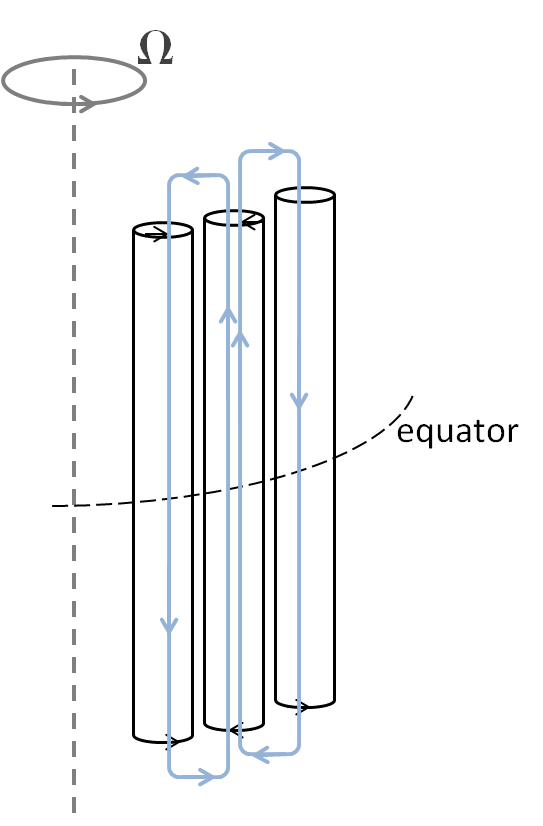}
	}
\caption{\small Diagram of columnar convection inside a rotating spherical shell.
\label{fig:column1}}
\end{center}
\end{figure}

Columnar convection consists of dominating nearly 2D (quasi-geostrophic) circular motions (i.e. independent of the $z$ cylindrical coordinate, defined by the rotation axis) with a secondary axial flow induced mainly by the boundaries along the columns for a Boussinesq fluid \citep[][]{Busse98,Olson99}. The 2D vortical motion arises from diverging up flow encountering the top and bottom boundaries, generating clockwise vortical motion. Since the resulting radial vorticity $\omega_r$ from the diverging up flow is negative in the North hemisphere and positive in the Southern (as a result of the Coriolis force acting on it), the $z$ component of the resulting vorticity $\omega_z$ is negative in both hemispheres. The flow will then sink toward the equatorial plane and converge to start rising again, generating now counter-clockwise motion \citep[positive $\omega_z$, see][and also Fig.~\ref{fig:column2}]{Olson99}. While the divergence of the flow occurs mainly near the boundaries, convergence can happen anywhere in the bulk and not exclusively at the equator (Fig.~\ref{fig:column2}). According to the Taylor-Proudman constraint \citep{Proudman16,Taylor22}, the local axes of vorticity tend to align with the rotation axis $z$, shaping columns of alternating vorticity sign. As illustrated in Fig.~\ref{fig:column1}, $u_z$ is positive in the northern hemisphere and negative in the southern along columns with positive $\omega_z$ and vice-versa. This means that in columnar convection, the kinetic helicity $H$ is dominated by the cylindrical component in $z$. Thus $H_z$ will always be negative in the northern hemisphere ($u_z$ and $\omega_z$ have opposite sign) and positive in the southern ($u_z$ and $\omega_z$ have the same sign). This helicity organization is helpful to obtain a large scale dipole field \citep[e.g][]{Olson99}. Note that columnar convection is also present in some plane-layer models (\citeauthor{Childress72} \citeyear{Childress72} and an example of application in numerical models, \citeauthor{Stellmach04} \citeyear{Stellmach04}), thus it does not necessarily depend on the spherical geometry.

A second regime, in which stratification and buoyancy instead play central roles, also often results in a similar kinetic helicity profile. In this ``plume-like" regime, more akin to the classic Rayleigh-Bénard convection problem, convection consists of rising and sinking flow between the inner and outer boundaries, predominantly along the radial direction. We generally expect that this type of convection will replace columnar convection in a rotating system when the effect of rotation is weak compared to buoyancy. 

This regime is easily attained near the surface of strongly stratified models of stellar and planetary convection (see, e.g., \citeauthor{Miesch08} \citeyear{Miesch08}; \citeauthor{Glatz09} \citeyear{Glatz09}; \citeauthor{Gastine13a} \citeyear{Gastine13a}; review in \citeauthor{Miesch09} \citeyear{Miesch09}), where the density gradient is much steeper leading to comparatively larger Rossby numbers \citep[\noc{ratio between inertia and Coriolis force, }e.g.][]{Browning08,Gastine12,Gastine13a}. Furthermore, a higher value of \noc{Rossby number} may also intensify helicity (see Sec.~\ref{models}\noc{ below}). According to the anelastic continuity equation,
\begin{equation}
	0 = \nabla\cdot (\rho\mathbf{u}) = 
		\underbrace{\rho\nabla\cdot\mathbf{u}}_\textrm{incompressible term}
		 + \underbrace{\mathbf{u}\cdot\nabla\rho}_\textrm{compressible term} \textrm{,}
\label{eq:anecont}
\end{equation}
when the density gradient is dominant (second term on the right side), a parcel rising along $r$ expands due to the rapidly decreasing density toward the surface (Fig.~\ref{fig:densprofs}). The Coriolis force acts on the expanding fluid to generate anticyclonic fluid motion \citep[negative $\omega_r$ in the northern hemisphere and positive in the southern, see][]{Glatz85}. In addition, the rising velocity decreases as a fluid parcel slows down when approaching the outer boundary, which may also decrease the effect of the incompressible term of the continuity equation (first term on the right side). For the same reason, sinking flow contracts generating cyclones (positive $\omega_r$ in the northern hemisphere and negative in the southern). Both effects of the rising/sinking flow naturally correlate as negative helicity, so plume-like convection near the surface typically gives a hemispherical North/South helicity pattern similar to that of columnar convection. Both types of convection often co-exist and reinforce each other in stratified models.

\subsection{``Inverted" helicity configuration}
\label{helicity2}

In the previous section, we described the most commonly observed case of helicity distribution in numerical dynamo models in rotating spherical shells. In the presence of milder density stratification ($N_\rho\!\lesssim\!4$) and rapid rotation, convection is often dominated by columnar convection. If on the other hand buoyancy dominates and the medium is highly stratified, plume-like convection (consisting of expanding upflows and contracting downflows) often prevails. Both lead to a kinetic helicity profile that is predominantly negative in the northern hemisphere. But what happens under other circumstances?

\begin{figure}
\begin{center}
{\centering
	\includegraphics[trim=0.5cm 0.0cm 0.5cm 0cm, height=1.9in]{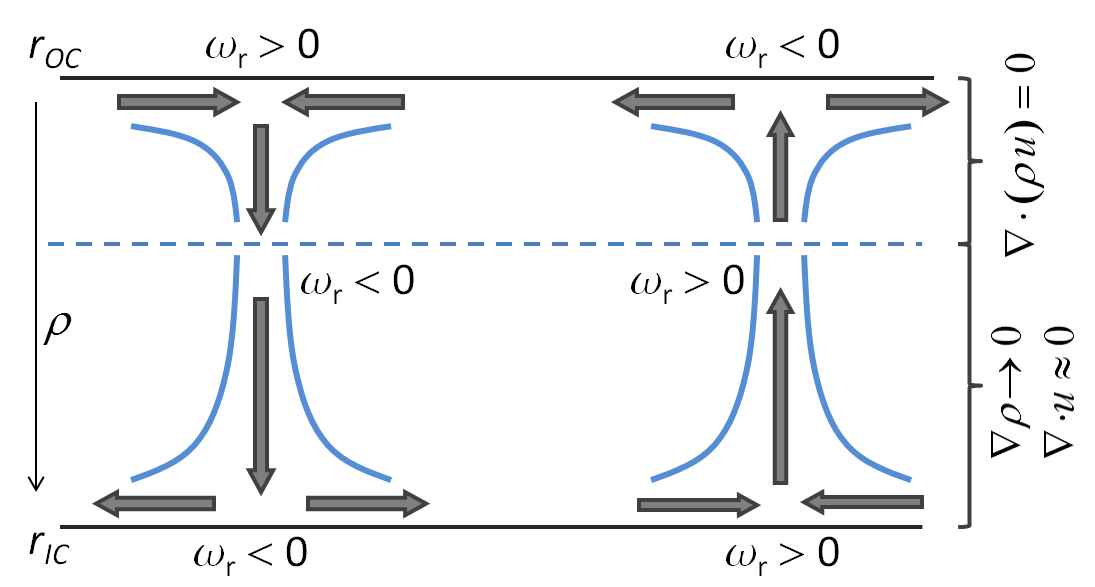}
	}
\caption{\small 2D portion of a spherical latitudinal section represented in a planar view. The dashed line separates the inner part of the density gradient where it is mild and the outer part where the gradient becomes steeper (see Fig.~\ref{fig:densprofs}). Note that this is merely a sketch, thus is it \noc{not} done to scale.
\label{fig:rbc}}
\end{center}
\end{figure}

In the presence of a weak density contrast, roughly $N_\rho\!\lesssim\!1$, a rising fluid parcel will tend to contract since its velocity is increasing as it starts from the bottom with $u_r\!=\!0$ (impenetrable boundaries). More generally, this may be true even far from the boundaries if $\partial u_r/\partial r\!>\!0$. This is easily understood from the equation of mass conservation (Eq.~\ref{eq:anecont}) when assuming that $\nabla\rho$ is small. Such small density contrasts typically exist in stratified models in the inner $80-90\%$ of the radius (see \noc{Sec.~\ref{mod} below, particularly Fig.~\ref{fig:densprofs}}). Consequently, the continuity equation becomes approximately the Boussinesq continuity equation of $\nabla\cdot\mathbf{u}\!=\!0$ in the deeper part of the shell. Figure~\ref{fig:rbc} illustrates schematically this behaviour below the dashed line which is a representation of the inner part of the radius of a spherical shell. In this region, the density gradient is relatively smaller than above the dashed line, which represents the outer few percent where typically most of the density gradient is located.

\noc{
In the region of small density gradient below the dashed line of Fig.~\ref{fig:rbc}, the Coriolis force acts on the rising+contracting fluid generating positive radial vorticity $\omega_r$ in the northern hemisphere and negative in the southern (cyclones). In the same region in the same way, sinking flow expands as it slows down toward the bottom boundary, generating anticyclones. In both cases of rising+contracting and sinking+expanding flows, the resulting sign of helicity is positive in the northern hemisphere and negative in the southern. This behaviour is opposite to the outer layer (described in the previous Section) and it seems only pertinent when columnar convection becomes minimal in the bulk.
}

In this schematic picture, explored in more detail in the following Sections, the inversion of kinetic helicity is a purely hydrodynamical effect, independent of the magnetic field, though it will ultimately be important to determine the direction of the poloidal field generated by the $\alpha$-effect and consequently the sign of the generated toroidal field, which explains the direction of propagation of a dynamo wave in the Parker-model.

It is also useful to distinguish between the possibility of deep layers where the kinetic helicity is ``reversed" (i.e., positive in the northern hemisphere), as explored here, from well-known boundary effects that would
tend to lead to the same profile within a narrow layer close to the bottom of the convection zone. It has long been anticipated that such reversals of kinetic helicity would arise at the base of the solar convection zone, for example, where convective downflows begin to be buoyantly braked and spread \citep[e.g.,][]{Brummell02}. But in simple parametrizations of the solar dynamo, the layer where this reversal is achieved is typically taken to be confined to a region near the bottom of the convective envelope or below its base \citep[e.g.,][]{Charbonneau10}. The convergence of flows to feed plumes that are beginning to rise buoyantly from a bottom thermal boundary layer is also well known \citep{Julien96,Nishikawa02,Dube13,Guervilly14} and is likewise somewhat distinct from the distributed helicity profiles described here, though the same basic physical mechanisms underlie all these. This will be demonstrated in the following Sections.

\section{Model}
\label{mod}

\subsection{Numerical model}
\label{anelasticeqs}

For this work, we used the code MagIC\footnote{https://github.com/magic-sph/magic} to solve an anelastic version of the MHD equations, following \cite{Glatz123}, \cite{Brag95} and \cite{Lantz99}, in a rotating spherical shell. This code implements a dimensionless formulation, where the length scale is the shell thickness $d$, the time is non-dimensionalized by the viscous time $\tau_\nu=d^2/\nu$ (where $\nu$ is the kinematic viscosity) and the temperature, gravity and density by their values at the outer boundary, respectively $T_o$, $g_o$ and $\rho_o$. Lastly, the magnetic field unit is given by $\sqrt{\Omega\mu\lambda_i\rho_o}$, where $\Omega$ is the rotation rate, $\mu$ the magnetic permeability and $\lambda_i$ the magnetic diffusivity at the bottom boundary of the domain, $r_i$. The dimensionless equations are
\begin{equation}
\begin{split}
	E\,\bigg(\frac{\partial \mathbf{u}}{\partial t} + \mathbf{u}\cdot\nabla \mathbf{u}\bigg)
	= &- \nabla\frac{p}{\tilde{\rho}} - 2\mathbf{e}_z\times\mathbf{u}
	+ \frac{Ra\,E}{Pr}\frac{r}{r_o}s\,\mathbf{e}_r
\\
	&+ \frac{1}{Pm_i\,\tilde{\rho}}(\nabla\times\mathbf{B})\times\mathbf{B}
	+ \frac{E}{\tilde{\rho}}\nabla\cdot\textsf{S} \textrm{,}
\end{split}
\label{eq:navierstokeseq}
\end{equation}
\begin{equation}
	\frac{\partial \mathbf{B}}{\partial t} = \nabla\times(\mathbf{u}\times\mathbf{B})
	- \frac{1}{Pm_i}\nabla\times(\tilde{\lambda}\nabla\times\mathbf{B}) \textrm{,}
\label{eq:inductioneq}
\end{equation}
\begin{equation}
\begin{split}
	\tilde{\rho}\,\tilde{T}\,\bigg(\frac{\partial s}{\partial t} + \mathbf{u}\cdot\nabla s\bigg)
		= &\frac{1}{Pr}\nabla\cdot (\tilde{\rho}\tilde{T}\nabla s)
			+ \epsilon\tilde{\rho}
\\
	&+ \frac{Pr\,Di}{Ra} \left[ Q_{\nu} + \frac{1}{Pm_i^2\,E} Q_j \right]
	\textrm{.}
\end{split}
\label{eq:energyeq}
\end{equation}
\begin{equation}
	\nabla\cdot (\tilde{\rho}\mathbf{u}) = 0 \textrm{,}
\label{eq:divfloweq}
\end{equation}
\begin{equation}
	\nabla\cdot\mathbf{B} = 0 \textrm{.}
\label{eq:divfieldeq}
\end{equation}
where \noc{$\mathbf{u}$, $\mathbf{B}$ and $s$ are the velocity, magnetic field and entropy fields, respectively,} $\eta$ is the aspect ratio given by $\eta\!=\!r_i/r_o\!=\!0.2$ and the traceless rate-of-strain tensor $\textsf{S}$ for homogeneous $\nu$ is
\begin{equation}
	\textsf{S} = 2\tilde{\rho}\bigg[\textsf{e}_{ij}-\frac{1}{3}\delta_{ij}\nabla\cdot\mathbf{u}\bigg] 
			\textrm{,   }\;\;\;\;
	\textsf{e}_{ij} = \frac{1}{2}\bigg(\frac{\partial u_i}{\partial x_j}+\frac{\partial u_j}{\partial x_i}\bigg) \textrm{,}
\label{eq:straintensor}
\end{equation}
and $\delta_{ij}$ is the identity matrix. The viscous and ohmic heating contributions are
\begin{equation}
	Q_{\nu}=2\tilde{\rho}\bigg[\textsf{e}_{ij}\textsf{e}_{ji}-\frac{1}{3}(\nabla\cdot\mathbf{u})^2\bigg]
\label{eq:vischeat}
\end{equation}
and
\begin{equation}
	Q_j=\tilde{\lambda}(\nabla\times\mathbf{B})^2 \textrm{.}
\label{eq:ohmicheat}
\end{equation}

\begin{figure}
\begin{center}
{\centering
	\includegraphics[trim=0cm 0.0cm 0.0cm 0cm, height=2.25in]{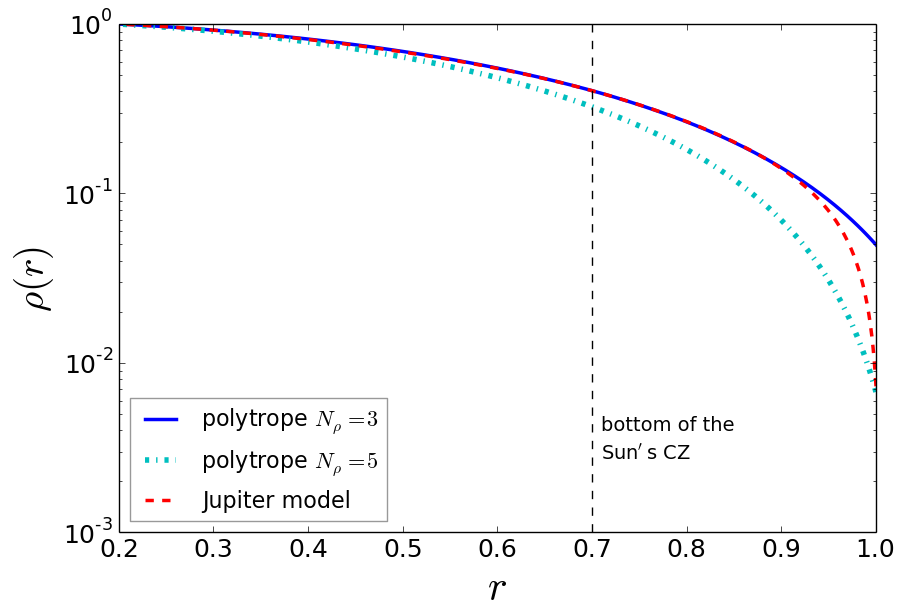}
	}
\caption{\small Background density profiles used in this work. The red dashed line
corresponds to Jupiter's density profile \citep{French12}, where $N_\rho\!\sim\!4.9$ in the inner $99\%$ of the shell radius and fitted by a polytrope of index $n\!\sim\!2.2$. The cyan dot-dashed line shows the background density profile of a polytrope with $N_\rho\!=\!5$ and $n\!=\!2$. The blue solid line shows the background density profile of a polytrope with $N_\rho\!=\!3$ and $n\!=\!2$.
\label{fig:densprofs}}
\end{center}
\end{figure}

The background reference state is defined by the temperature gradient $d\tilde{T}/dr=-Di\,g(r)$ and the background density is derived as $\tilde{\rho}(r)=\tilde{T}^n$\noc{, where the tilde corresponds to the background reference state} and $n$ is the polytropic index. \noc{The dissipation number $Di$ is expressed as}
\begin{equation}
	Di = \frac{g_o\,d}{c_p\,T_o} = 2\,\frac{\textrm{e}^{N_\rho/n}-1}{1+\eta}
	\textrm{,}
\label{eq:dissnum}
\end{equation}
where $N_\rho=\ln{(r_i/r_o)}$ is the number of density scale heights and $r_i$ and $r_o$ are the radii of the inner and outer boundaries\noc{, respectively}. In Fig.~\ref{fig:densprofs}, the solid blue line corresponds to $N_\rho\!=\!3$ and $n\!=\!2$ and the cyan dot-dashed line to $N_\rho\!=\!5$ and $n\!=\!2$. The red dashed line corresponds to a polynomial fit of Jupiter's interior model \citep{French12}, where the corresponding $\tilde{T}(r)$ was obtained from the best corresponding $n$ to the data for the background temperature from the same model. The result is $N_\rho\!\approx\!4.9$ and $n\!\approx\!2.2$. 

Finally, the control parameters in Eqs.~(\ref{eq:navierstokeseq}--\ref{eq:divfieldeq}) correspond to ratios between pairs of terms of the Navier-Stokes equation, namely the Ekman number (viscous over Coriolis force), the Rayleigh number (measure of the convective driving of the system) and the fluid and magnetic Prandtl numbers,
\begin{equation}
	E=\frac{\nu}{\Omega d^2} \textrm{,}
\label{eq:ekmunber}
\end{equation}
\begin{equation}
	Pr=\frac{\nu}{\kappa} \textrm{,}
\label{eq:prmunber}
\end{equation}
\begin{equation}
	Pm_i=\frac{\nu}{\lambda_i} \textrm{,}
\label{eq:pmmunber}
\end{equation}
where $\kappa$, $\nu$ and $\lambda$ are the thermal, viscous and magnetic diffusivities, respectively. In our models, we fixed $E\!=\!10^{-4}$ and we used $Pr\!=\!0.1/1.0$ and $Pm_i\!=\!0.0-20.0$ for the ratios between diffusivities. \noc{The Rayleigh number is either entropy-based or flux-based as follows
\begin{equation}
	Ra=\frac{g_o d^3}{\nu\kappa}\,S_{scale} \textrm{,}
\label{eq:ramunber}
\end{equation}
where $S_{scale}=\Delta s/c_p$ if the boundary conditions of the energy equation are fixed entropy and $S_{scale}=q_o d/(\rho_o c_p\kappa)$ for the models where the boundaries are assumed fixed flux. The quantities $q_o$ and $c_p$ are the specific entropy flux at the outer boundary and the specific heat capacity at constant pressure, respectively.
}

Even though the results discussed in this paper were first found in our Jupiter models with variable magnetic diffusivity \citep[as reported by][]{Jones14}, we simplified some of these models by carrying out a small number of hydrodynamic and magnetic simulations with constant transport properties along radius, to illustrate our conclusions. The variable conductivity profile used in the Jupiter models is described in \cite{Gastine14}.

\subsection{Numerical method}
\label{code}

The MagIC code uses a pseudo-spectral method to solve the MHD equations described above. Spherical harmonics are used in the horizontal direction ($\theta,\phi$) up to degree and order $\ell_{max}$ and Chebyshev polynomials in the radial direction up to degree $N_r$ \citep[see][for a more detailed description]{Wicht02}. The equations are solved in the commonly used poloidal/toroidal decomposition of the divergence-free fields $\rho\mathbf{u}$ and $\mathbf{B}$ (Eqs.~\ref{eq:divfloweq} and \ref{eq:divfieldeq}) as
\begin{equation}
\begin{split}
	&\rho\mathbf{u} = \nabla\times\nabla\times v\,\mathbf{e}_r + 
					\nabla\times w\,\mathbf{e}_r
\\
	&\mathbf{B} = \nabla\times\nabla\times b\,\mathbf{e}_r + 
					\nabla\times t\,\mathbf{e}_r \textrm{,}
\label{eq:poltor}
\end{split}
\end{equation}
where $v$ and $b$ are the poloidal potentials of $\rho\mathbf{u}$ and $\mathbf{B}$, respectively, while $w$ and $t$ are the toroidal counterparts. The radial unit vector is represented by $\mathbf{e}_r$.

The boundary conditions applied in our simulations are the same for the velocity and magnetic fields. Following previous work which had the purpose of modelling the gas giants \citep{Gastine12,Gastine12a,Duarte13}, the inner boundary is assumed no-slip and the inner core is modelled as an electrical conductor. Top boundary is considered free-slip and the magnetic field is there matched to a potential field. The entropy boundary conditions and heating modes vary in our models. Several of our cases assume simple bottom heating, i.e. the entropy contrast between the bottom and top boundaries $\Delta s$ is fixed and there is no internal heating in the system\noc{, thus the heating entering the system through the bottom boundary exits through the top. The other two set-ups for the energy equation considered here account for internal heat sources: in one case with fixed entropy boundaries and in the other with fixed flux boundary conditions.}

\subsection{Diagnostic parameters}
\label{corr}

In the following Sections we will describe the flow by often referring to several dimensionless diagnostic parameters to be consistent with the formulation outlined in the previous Sections.

The amplitude of the flow contributions is measured in terms of the Rossby numbers $Ro$. The value of $Ro$ is calculated as
\begin{equation}
	Ro=\frac{u}{\Omega\,d}
	\textrm{,}\;\;\textrm{ with }\;\;
	u=\sqrt{\frac{3}{r_o^3-r_i^3}\int_{r_i}^{r_o}\!\langle u^2\rangle\,r^2\,\mathrm{d}r}
	\,\textrm{,}
\label{eq:rozon}
\end{equation}
where $u$ is the rms volume-averaged flow velocity and the triangular brackets denote the angular average
\begin{equation}
	\big\langle f\big\rangle=\frac{1}{4\pi} \, \int_{0}^{\pi}\! \int_{0}^{2\pi}\!
		f(r,\theta,\phi) \sin\!\theta \, \mathrm{d}\theta \, \mathrm{d}\phi
	\textrm{.}
\label{eq:volave}
\end{equation}

The local Rossby number has been introduced by \cite{Christensen06} to quantify the relative importance of the advection term in the Navier-Stokes equation (Eq.~\ref{eq:navierstokeseq}). Here we consider only the non-axisymmetric part of the flow velocity $u_{m\neq0}$ (i.e. velocity excluding the axisymmetric flow component) to calculate the local convective Rossby number, defined as
\begin{equation}
	Ro_{\ell\,conv}=
		\frac{\sqrt{\frac{1}{V}\,\int_{r_i}^{r_o}\! \big\langle u_{m\neq0}^2 \big\rangle
		\,r^2\, \mathrm{d}r}}{\Omega\,\ell}
	\textrm{,}
\label{eq:rol}
\end{equation}
where, $\ell$ is a typical flow length scale given by
\begin{equation}
	\ell(r)=\frac{\pi \, u^2(r)}{\displaystyle\sum\limits_l l \, u_l^2(r)}
	\textrm{.}
\label{eq:ell}
\end{equation}
Here $u_l$ is the flow contribution of spherical harmonic degree $l$.

The geometry of the surface field is characterized by the dipolarity
\begin{equation}
	f_{dip}=\frac{\Big\langle \big({\mathbf{B}_{l=1}^{m=0}}\big)^2 \Big\rangle}
		{\Bigg\langle \displaystyle\sum\limits_{l,m\le l_{max}} \big({\mathbf{B}_l^m}\big)^2
		\Bigg\rangle}
	\textrm{,}
\label{eq:dip}
\end{equation}
which measures the relative energy in the axial dipole contribution at the outer boundary $r_o$.

Finally, to describe the behaviour of the various components of the flow, we used correlations between pairs of variables. A correlation between two sets of data $A$ and $B$ in the form of 3D matrices corresponding to the three spherical coordinates $r,\theta,\phi$ is calculated at each radial level as
\begin{equation}
\begin{split}
	& \textrm{corr}(A,B)_{\theta,\phi}(r) =
\\
		&= \frac{ \sum_{\theta,\phi}
		\left[ A_{\theta,\phi}(r) - \langle A_{\theta,\phi}(r) \rangle\right]
			\left[ B_{\theta,\phi}(r) - \langle B_{\theta,\phi}(r) \rangle\right]
		}{
		\sqrt{
		\displaystyle\sum_{\theta,\phi} \left[A_{\theta,\phi}(r) - \langle A_{\theta,\phi}(r) \rangle \right]^2
		\displaystyle\sum_{\theta,\phi} \left[B_{\theta,\phi}(r) - \langle B_{\theta,\phi}(r) \rangle \right]^2 }
		} \textrm{.}
\label{eq:corr}
\end{split}
\end{equation}

The parameters described in this Section are listed along control parameters in Tab.~\ref{TabRes}, averaged over at least $0.1$ viscous time. The values of the Rayleigh number necessary for the onset of convection as defined by \cite{Jones09a} (i.e., where all other dimensionless control parameters are fixed), were obtained for cases without internal heating. These values were used to obtain the supercriticality values listed in Tab.~\ref{TabRes} at constant Ekman number. In the same table, simulations without internal heating correspond \noc{to $\epsilon\!=\!0$ (see Eq.~\ref{eq:energyeq}) and $\epsilon\!\neq\!0$ if internal heating is present}. Concerning the entropy boundary conditions corresponding to columns $s_{BC,i}$ (bottom) and $s_{BC,o}$ (top), values $0$ and $1$ indicate fixed entropy and flux, respectively.

\section{Kinetic helicity realized in non-linear simulations}
\label{results1}

\subsection{Numerical hydro/dynamo models}
\label{models}

\begin{figure}
\begin{center}
{\centering
	\includegraphics[trim=0cm 0cm 0.5cm 0cm, height=2.4in]{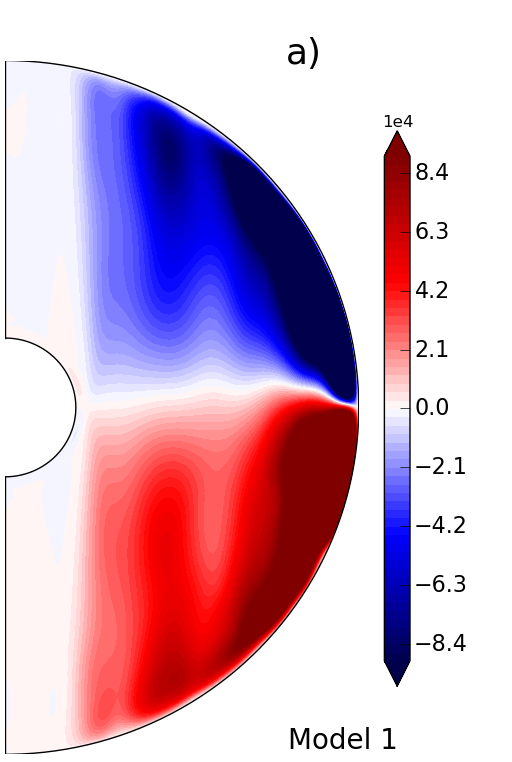}
	\includegraphics[trim=0cm 0cm 0.5cm 0cm, height=2.4in]{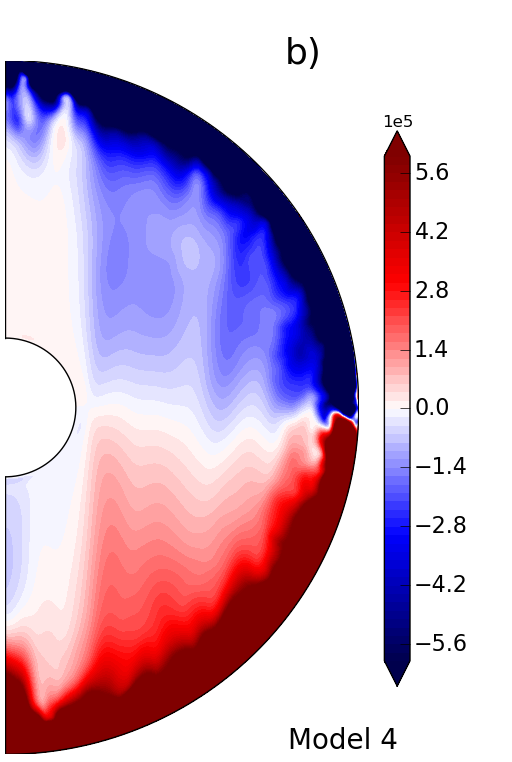}
	\includegraphics[trim=0cm 0cm 0.5cm 0cm, height=2.4in]{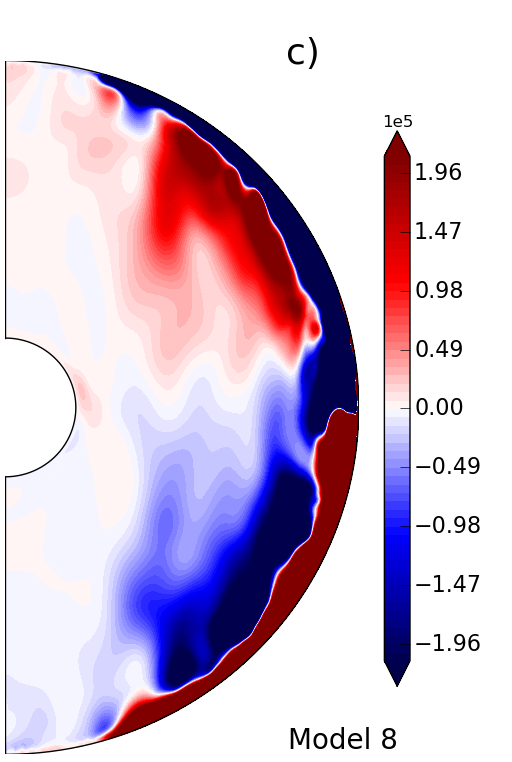}
	\includegraphics[trim=0cm 0cm 0.5cm 0cm, height=2.4in]{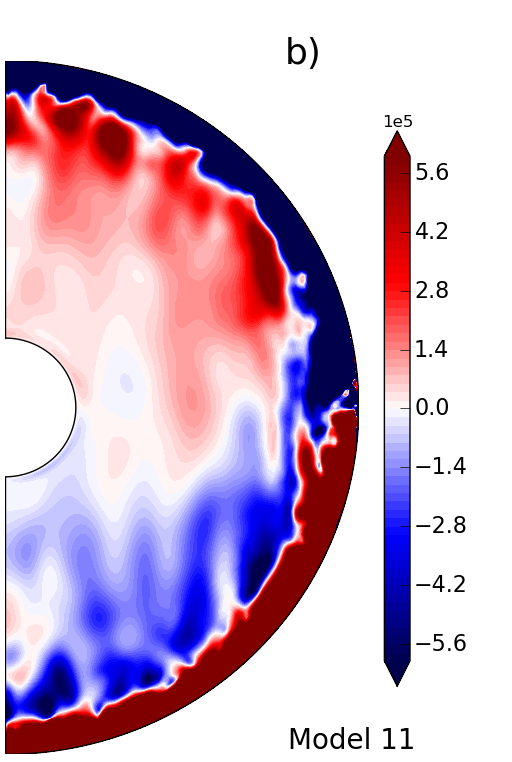}
}
\caption{\small Azimuthally-averaged contour plots of kinetic helicity. The top row displays the helicity pattern described in Sec.~\ref{helicity} for models $1$ ($\mathbf{a}$) and $4$ ($\mathbf{b}$) of Tab.~\ref{TabRes}. The bottom row shows two examples of the regime described in Sec.~\ref{helicity2} for cases $8$ ($\mathbf{c}$) and $11$ ($\mathbf{d}$) of the same Table.
\label{fig:kinhel}}
\end{center}
\end{figure}

\begin{figure}
\begin{center}
{\centering
	\includegraphics[trim=0cm 0cm 0.5cm 0cm, height=2.4in]{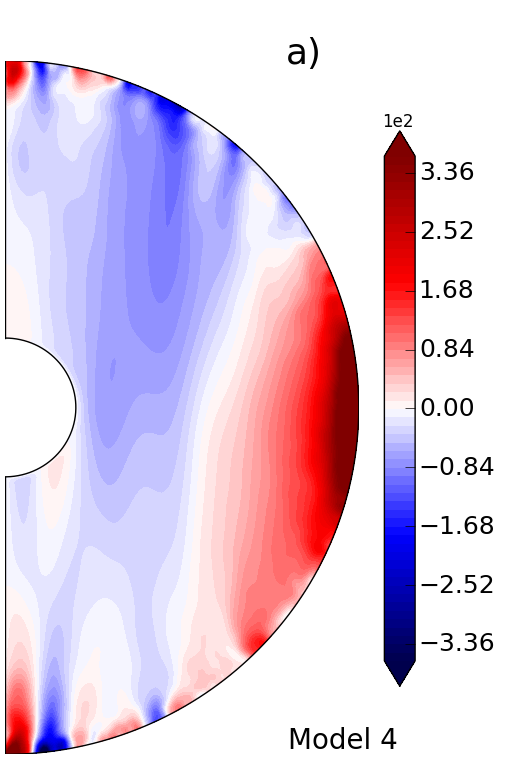}
	\includegraphics[trim=0cm 0cm 0.5cm 0cm, height=2.4in]{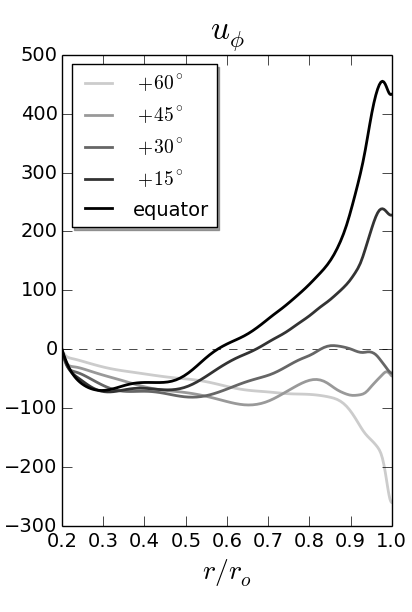}
	\includegraphics[trim=0cm 0cm 0.5cm 0cm, height=2.4in]{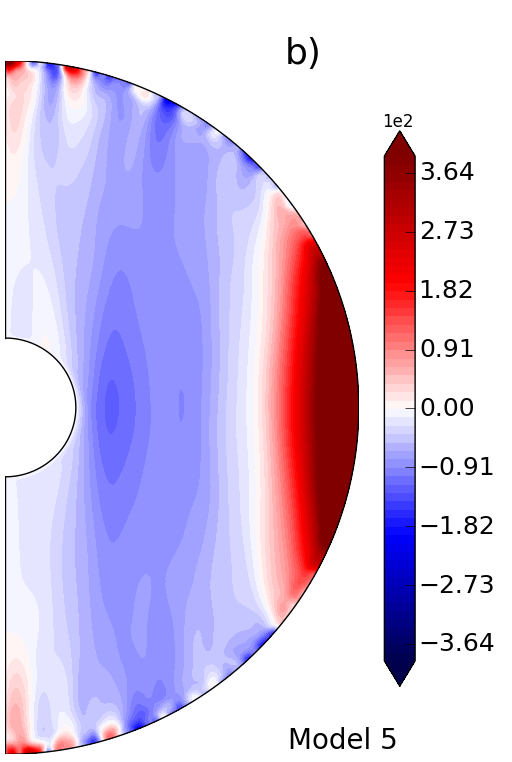}
	\includegraphics[trim=0cm 0.5cm 0.5cm -2cm, height=2.75in]{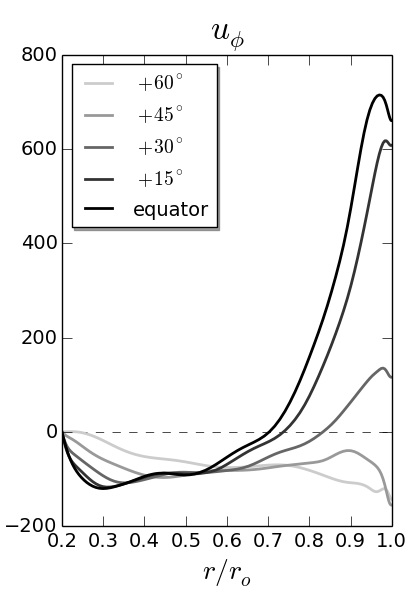}
	\includegraphics[trim=0cm 0cm 0.5cm 0cm, height=2.4in]{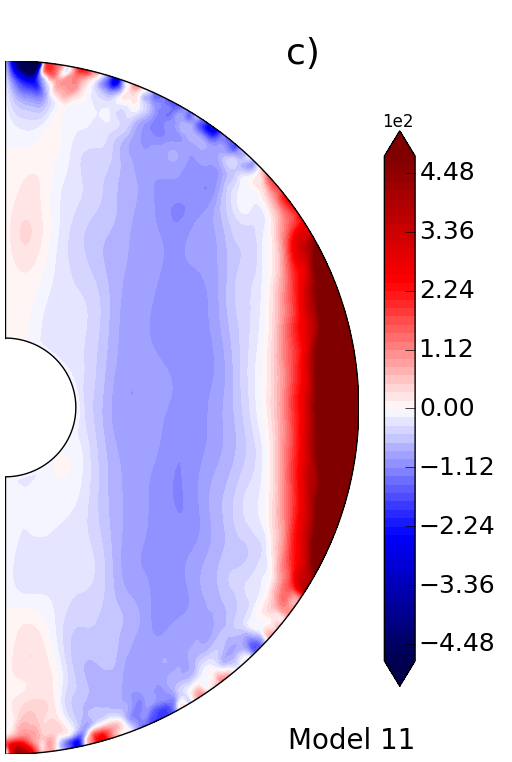}
	\includegraphics[trim=0cm 0.5cm 0.5cm -2cm, height=2.75in]{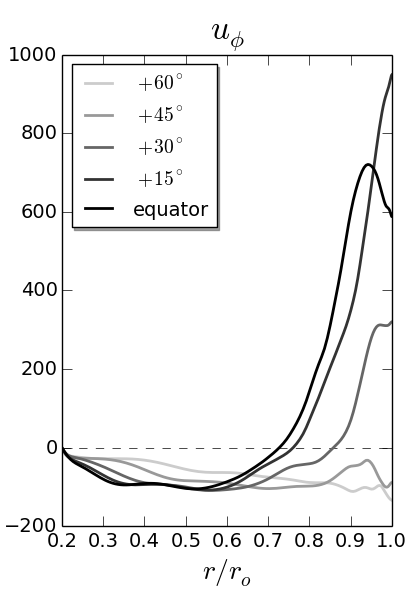}
}
\caption{\small Azimuthally-averaged contour plots of zonal velocity $u_\phi$. The top row displays \noc{model $4$ from Tab.~\ref{TabRes}, the middle row model $5$ and the bottom row model $11$. The two top rows are examples of the helicity pattern described in Sec.~\ref{helicity} and the bottom row of Sec.~\ref{helicity2}}. The velocity is given in units of Reynolds number, $Ro/E$ (see Eq.~\ref{eq:rozon}).
\label{fig:vphi}}
\end{center}
\end{figure}

Figure~\ref{fig:kinhel} illustrates the two helicity configurations described in Secs.~\ref{helicity} and Sec.~\ref{helicity2} for global hydrodynamical and dynamo simulations in a rotating spherical shell, for two different supercriticalities. Contour plots of azimuthally-averaged kinetic helicity are shown here for snapshots, since these helicity patterns are not time-dependent in our models. The top panels show the usual one-layer negative(North)/positive(South) helicity configuration for the combination of columnar convection and plume-like convection from Sec.~\ref{helicity}. The bottom plots show the two-layer behaviour described in Sec.~\ref{helicity2} and illustrated in Fig.~\ref{fig:rbc}. In each row, the supercriticality of the models increases from left to right, hence the later onset of convection inside the tangent cylinder TC (cylinder tangent to the inner core boundary around the rotation axis) on the left-side panels. As mentioned in the Introduction, the different sign of helicity (e.g. top and bottom rows of Figure~\ref{fig:kinhel}) is obtained for models which retain similar differential rotation profiles, as shown in Figure~\ref{fig:vphi}. \noc{As evident in Fig. 6, the differential rotation is solar-like in the sense of having a faster equator. Model 4 does however exhibit some non-geostrophic flow that breaks the equatorial symmetry \citep{Gastine12a}. This Figure also shows an additional model (case $5$ of Tab.~\ref{TabRes}) because we refer to that simulation later in Sec.~\ref{waves}.}

\begin{figure}
\begin{center}
{\centering
	\includegraphics[trim=0.1cm 0cm 0.5cm 0cm, height=1.73in]{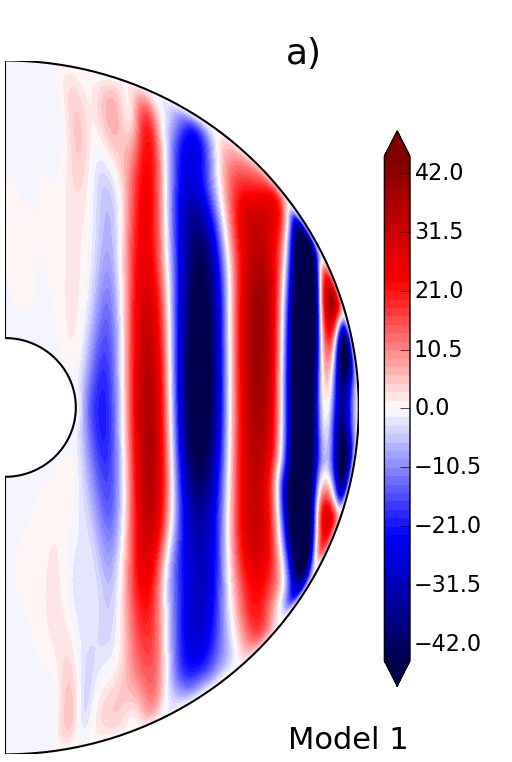}
	\includegraphics[trim=0.4cm 0cm 0.5cm 0cm, height=1.73in]{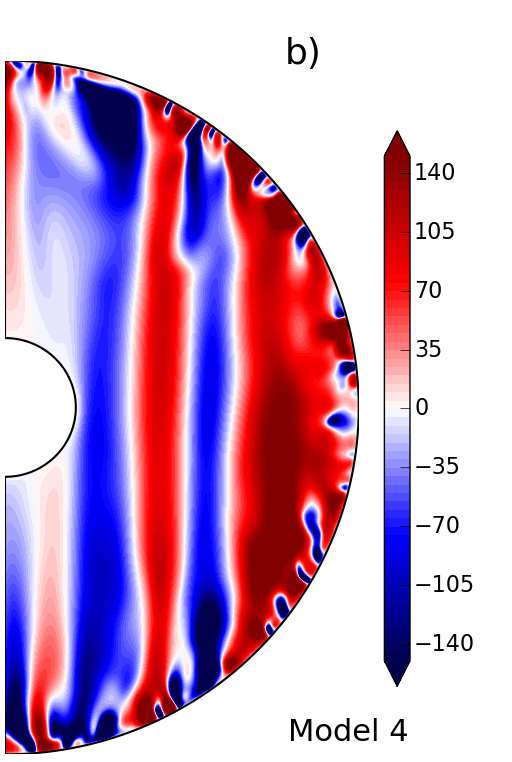}
	\includegraphics[trim=0.8cm 0cm 0.5cm 0cm, height=1.73in]{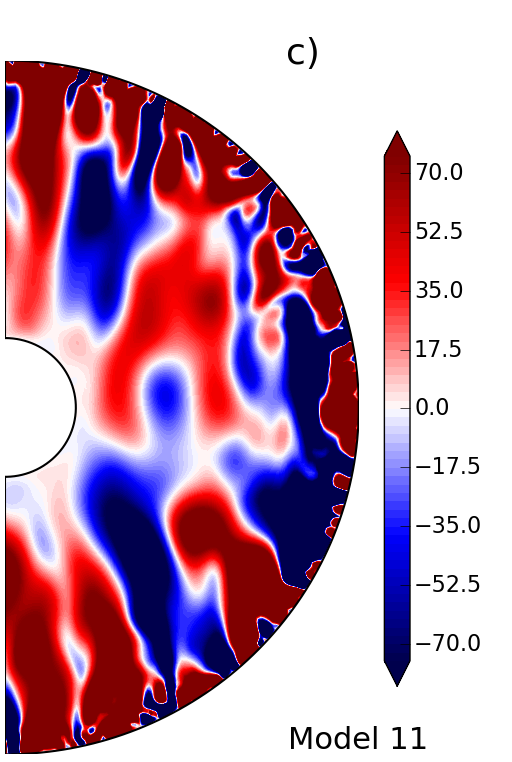}
}
\caption{\small Contour plots of slices of radial velocity for three of the cases shown in Fig.~\ref{fig:kinhel}, namely cases $1$, $4$ and $11$ of Tab.~\ref{TabRes} from left to right, corresponding to panels $\mathbf{a}$, $\mathbf{b}$ and $\mathbf{c}$, respectively. The velocity is given in units of Reynolds number, $Ro/E$ (see Eq.~\ref{eq:rozon}).
\label{fig:vrax}}
\end{center}
\end{figure}

\begin{figure*}
\begin{center}
{\centering
	\includegraphics[trim=0.2cm 0cm 1.0cm 0cm, height=1.5in]{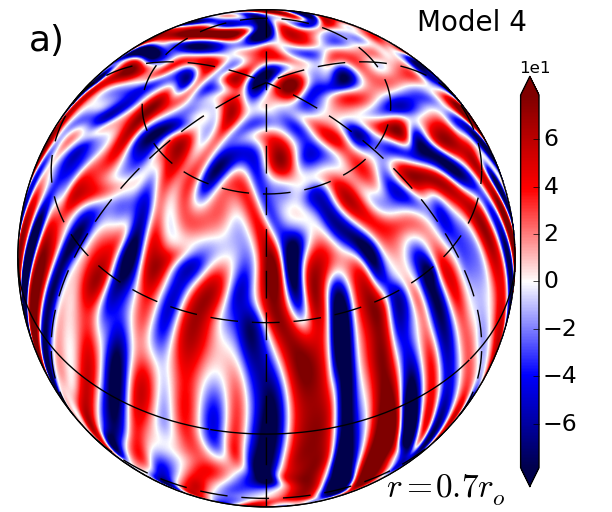}
	\includegraphics[trim=0.2cm 0cm 0.5cm 0cm, height=1.5in]{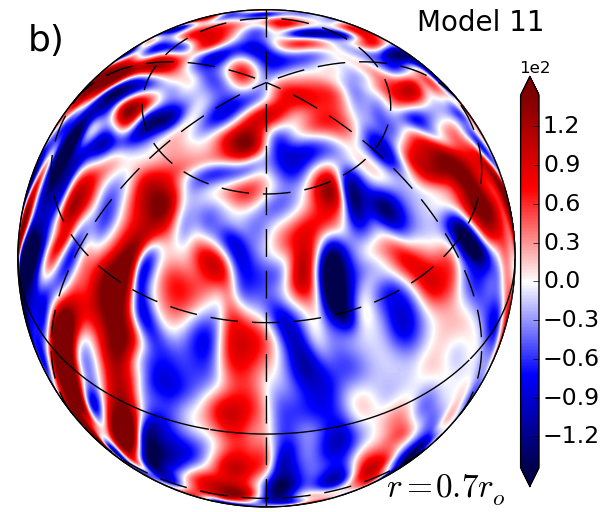}
	\includegraphics[trim=0.2cm 0cm 0.5cm 0cm, height=1.5in]{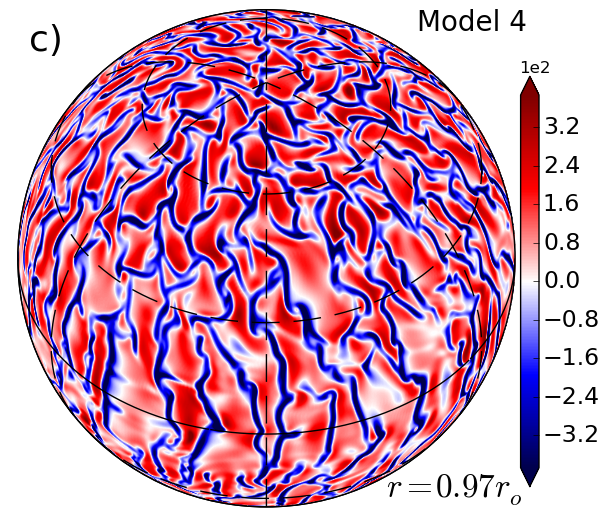}
	\includegraphics[trim=0.2cm 0cm 0.5cm 0cm, height=1.5in]{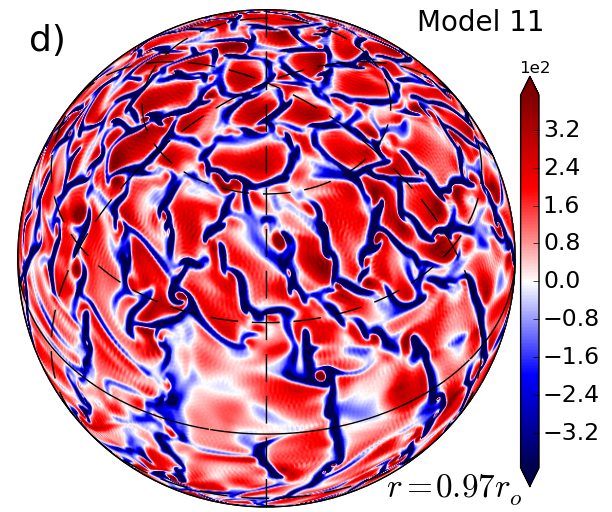}
}
\caption{\small Contour plots of radial velocity $u_r$ for the two models on the right column of Fig.~\ref{fig:kinhel}, namely models $4$ (panels $\mathbf{a}$ and $\mathbf{c}$) and $11$ (panels $\mathbf{b}$ and $\mathbf{d}$) of Tab.~\ref{TabRes}. Panels $\mathbf{a}$ and $\mathbf{b}$ correspond to the inner radial level of $r\!=\!0.7r_o$ and $\mathbf{c}$ and $\mathbf{d}$ to $r\!=\!0.97r_o$. The latitude circles and meridians (black dashed lines) are placed $60^\circ$ apart. The velocity is given in units of Reynolds number, $Ro/E$ (see Eq.~\ref{eq:rozon}).
\label{fig:vrax3d}}
\end{center}
\end{figure*}

Figure~\ref{fig:vrax} shows radial velocity contour plots for three cases displayed in Fig.~\ref{fig:kinhel}, namely the top row and the right panel of the bottom row\noc{. Figure~\ref{fig:vrax3d} displays models $4$ and $11$ in orthographic projection to illustrate the latitudinal/longitudinal distribution of the flow features at the two depths described in Secs.~\ref{helicity} and \ref{helicity2}}. The leftmost panel of Fig.~\ref{fig:vrax} (and top left panel of \ref{fig:kinhel}) is distinctly dominated by columnar convection. The middle panel has a stronger density gradient in the outer part of the shell as Fig.~\ref{fig:densprofs} shows, resulting in a break down of the columns in the outer radius, where convection becomes dominated by radial features \citep[see Fig. 12 of][]{Gastine13a}. Below this outer layer, convection remains under a columnar regime, though combined with plume-like convection. This set-up corresponds to the one-layer helicity pattern described in Sec.~\ref{helicity}, where both types of convection co-exist in strongly stratified models. The right panel corresponds to the two-layer helicity pattern described in Sec.~\ref{helicity2}, where convection is seemingly not columnar any more. In this case, the density gradient used was the interior model for Jupiter in Fig~\ref{fig:densprofs}, which may help achieving this configuration since, even though the density contrast is also around 5 density scale heights, it now concentrates most of the gradient in the outer $10-20\%$ of the radius of the shell. However, this is not the only difference, as we will discuss next.

Perhaps more significantly, the cases in Figs.~\ref{fig:kinhel}--\ref{fig:vrax3d} that show ``inverted" helicity (models $8$ and $11$) also differ from the other cases displayed here in these Figures by adopting a lower value of $Pr$ and a different mode of heating. While $Pr\!=\!1.0$ in the top row panels of Fig.~\ref{fig:kinhel}, as in most of our previous work \citep[for example,][]{Gastine12,Gastine12a,Yadav13a,Duarte13}, the cases in the bottom row have a lower value of $Pr\!=\!0.1$. The effect of lowering the Prandtl number has been thoroughly studied by \cite{Simitev05,Sreenivasan06}, where their main conclusion was the significant increase of the role of inertia when decreasing $Pr$ by one order of magnitude or more from unity. When inertia enters the force balance at a significant degree, the columnar convection constraint weakens in the bulk, as discussed by \cite{Christensen06}. The change of heating mode \noc{in conjunction with fixed flux boundaries} is also crucial\noc{. An internal heating source} has been shown to spread convection in the domain, thus also weakening convection in the bulk as it tends to detach the deeper convective motions from the inner boundary \citep{Hori10}.\noc{ With fixed entropy boundary conditions, on the other hand, changes in the internal heating mode did not appear to change the outcome in our simulations \citep[see however][for a possible counter-example]{Jones14}.}

\begin{figure}
\begin{center}
{\centering
	\includegraphics[trim=0cm 0.0cm 0.5cm 0cm, height=2.5in]{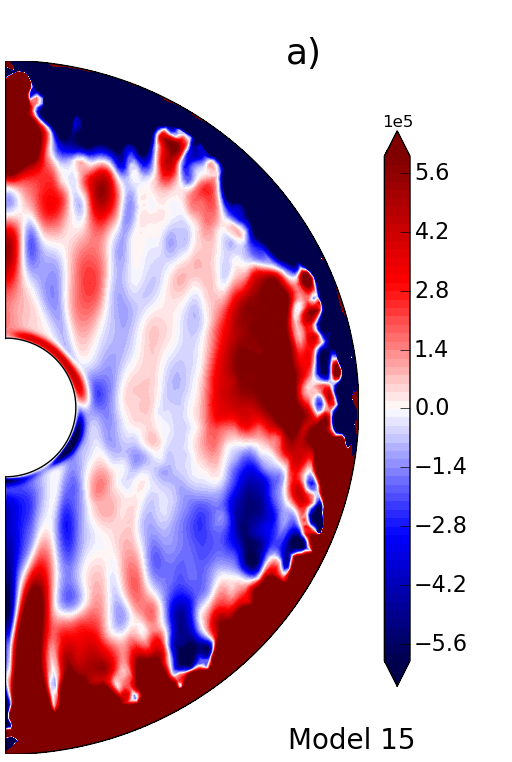}
	\includegraphics[trim=0cm 0.0cm 0.5cm 0cm, height=2.5in]{Helax_vcondInteriorModelRa4e9Pm2Pr01bchRaf.png}
}
\caption{\small Azimuthally-averaged countour plots of kinetic helicity for 2 cases with $Pr\!=\!0.1$. The heating mode of the model in panel $\mathbf{a}$ is bottom heating (model $15$ from Tab.~\ref{TabRes}) and on $\mathbf{b}$ is internal heating (model $11$ from Tab.~\ref{TabRes}).
\label{fig:pr}}
\end{center}
\end{figure}

Lowering the value of $Pr$ appears to be the most important factor in changing the helicity pattern due to its relation to the amount of inertia in the system, with a tendency to promote plume-like convection. When applying only a different heating mode or a different density profile or both, we saw very little difference in the final results of our models as the one-layer helicity configuration remains dominant. However, lowering $Pr$ alone is not sufficient either. In Fig.~\ref{fig:pr}, we show two cases with the same lower Prandtl number $Pr\!=\!0.1$, though convection in the left panel is driven by bottom heating and on right by internal heating proportional to the background density profile \noc{(Eq.~\ref{eq:energyeq})}. The left panel shows that indeed the inversion to attain the helicity regime of Sec.~\ref{helicity2} is not complete unless we combine the lower $Pr$ with internal heating.

In conclusion, the combination of the three effects (high $N_\rho$, low $Pr$ and internal heating) is important to achieve the required degree of non-columnar/plume-like convection in the bulk of the shell. The requirement of high $N_\rho$ is harder to constrain and appears to be inconsistent with the requirement for negligible density contrast in the interior. This apparent contradiction is because the higher the overall density contrast, the closer we get to a negligible density gradient in the bulk.  Even though this does not eliminate the possibility of Sec.~\ref{helicity2} also occurring in a weakly stratified or even Boussinesq model, cases \noc{$31$} and \noc{$32$} in Tab.~\ref{TabRes} show a helicity pattern very similar to the plane layer configuration, with a symmetry about half of the height \citep{Julien96}, or half the radius in the case of a spherical shell. How exactly this symmetry about the mid-plane is broken by the combination of density stratification and internal heating, and whether the ``inverted" helicity layers studied here require both these elements or are conceptually unrelated, requires further study.

\cite{Kapyla13} has reported a relation between the rotation rate and the degree of stratification for the onset of oscillatory solutions, which we did not explore here: they find that at higher $N_\rho$, lower $Ro$ is necessary to find oscillatory solutions. They also argued that equatorward wave solutions occur only at larger $N_\rho$, whereas poleward propagation is found instead at milder density contrasts.

\subsection{Dynamo waves}
\label{waves}

\begin{figure*}
\begin{center}
{\centering
	\includegraphics[trim=0cm 0cm 0.5cm 0cm, height=2.4in]{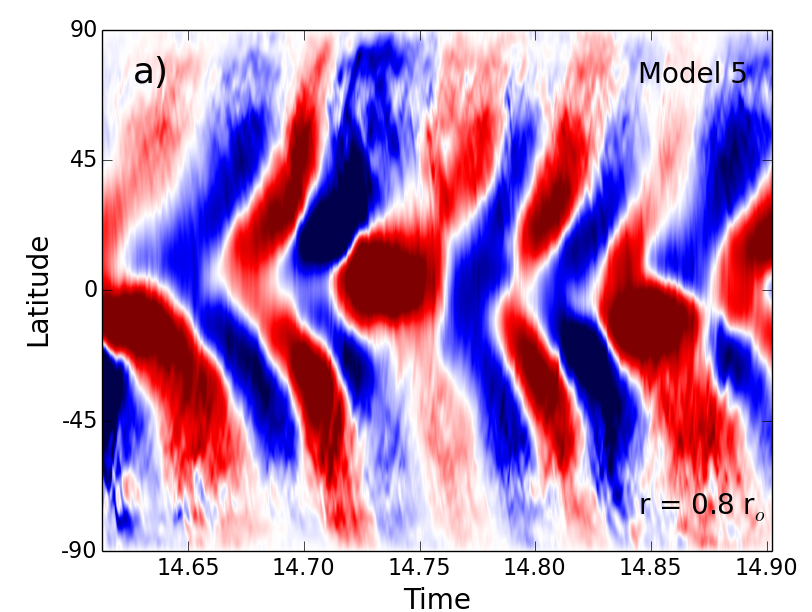}
	\includegraphics[trim=0cm 0cm 0.5cm 0cm, height=2.4in]{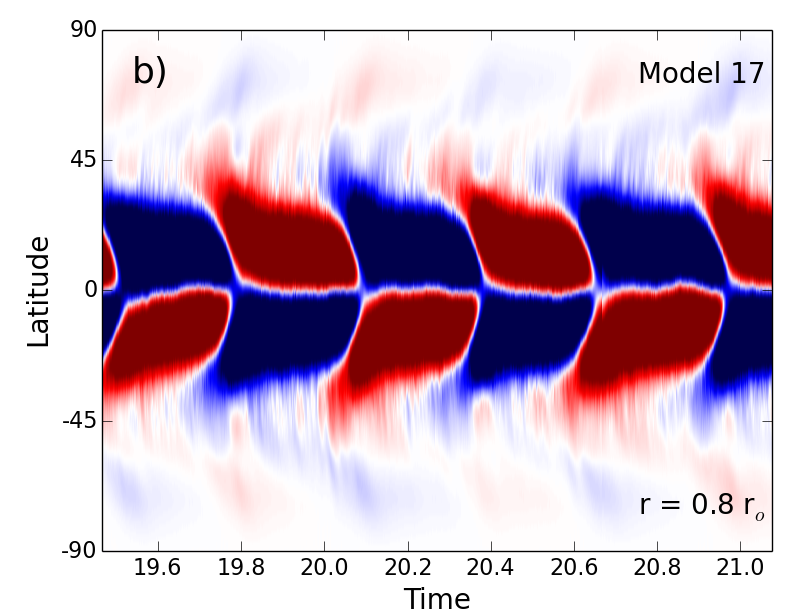}
	\includegraphics[trim=0cm 0cm 0.5cm 0cm, height=2.4in]{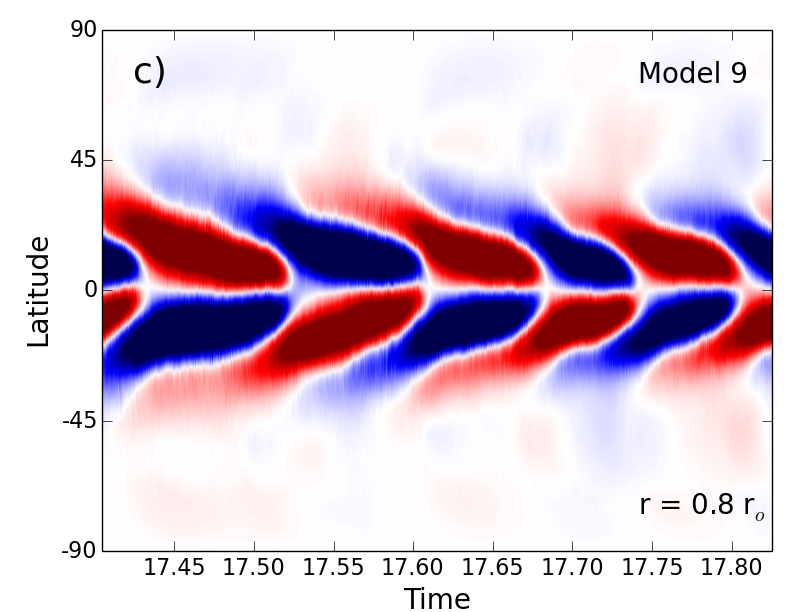}
	\includegraphics[trim=0cm 0cm 0.5cm 0cm, height=2.4in]{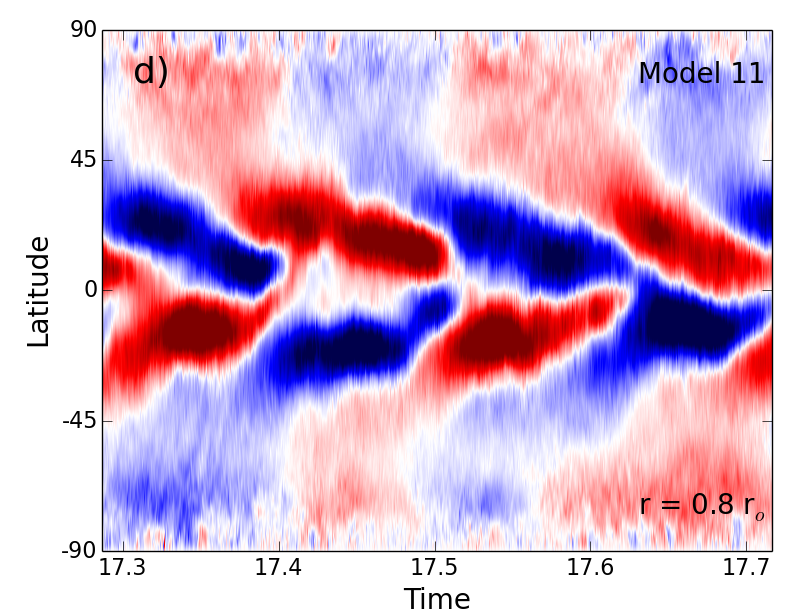}
}
\caption{\small Four examples of models from our database that illustrate the effect of the helicity regime in the direction of propagation of the dynamo wave. The time-evolution of the toroidal component of the magnetic field is shown averaged over longitude. The time is given in viscous time $\tau_\nu$. The cases in correspond to models $5$, $17$, $9$ and $11$ of Tab.~\ref{TabRes} in clockwise direction, starting from the top left panel.
\label{fig:waves1}}
\end{center}
\end{figure*}

In Fig.~\ref{fig:waves1} we show a few examples of butterfly diagrams constructed from our models, to illustrate the consequence of the different helicity patterns described above, in rough accordance with the Parker model. The top left panel shows a previous simulation from our previous Jupiter models which has a poleward-propagating dynamo wave for comparison \citep[see][]{Gastine12a,Duarte13}. The other three panels exemplify the variety of equatorward propagating waves found in the models described here. In some cases, the equatorward part is confined to a lower latitudinal band with weak poleward propagation at high latitudes, likely due to the lack of convection inside the TC at lower supercriticality (see examples of Fig.~\ref{fig:kinhel}). Nonetheless, this higher-latitude feature is common in most models and a similar higher latitude feature has also been observed in the Sun \citep{Hathaway14}.

\begin{figure*}
\begin{center}
{\centering
	\includegraphics[trim=0cm 0cm 0.5cm 0cm, height=2.45in]{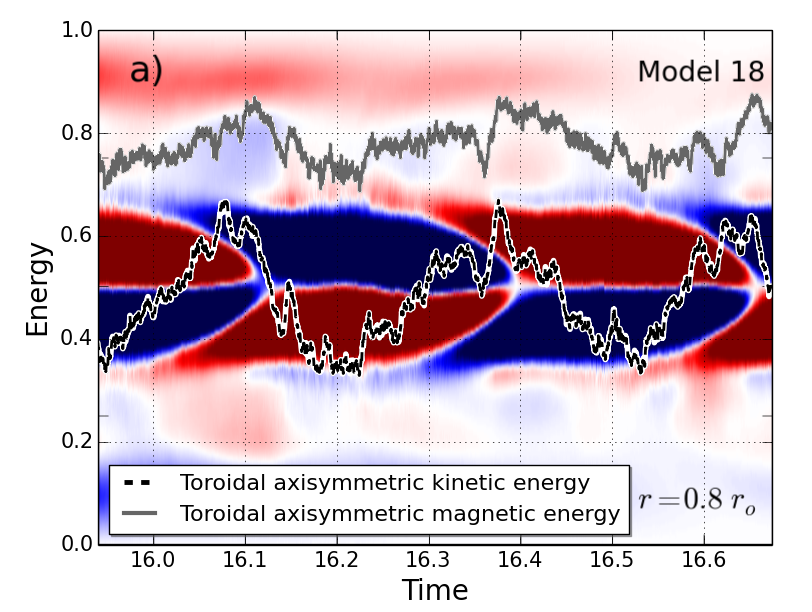}
	\includegraphics[trim=0cm 0cm 0.5cm 0cm, height=2.45in]{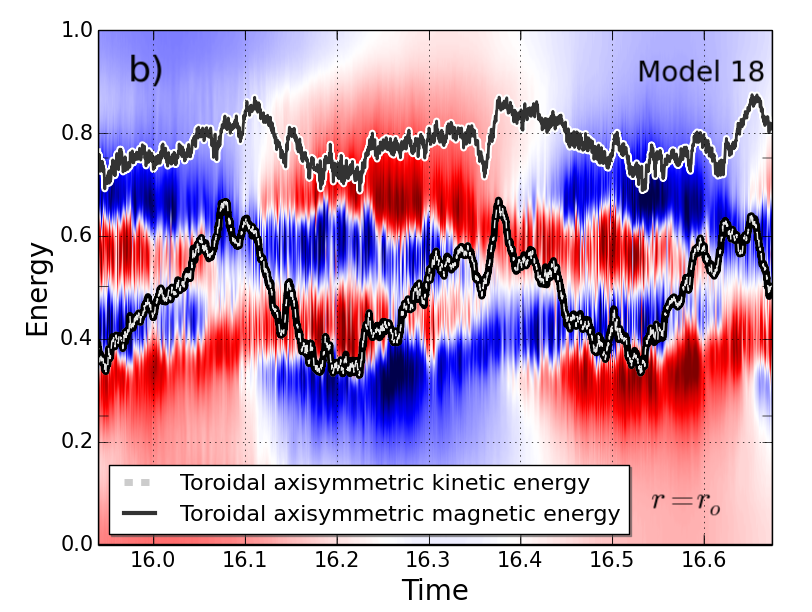}
}
\caption{\small Illustration of the wave at $80\%$ of the radius (left panel) and the resulting signature in the surface field (right panel), for model $18$ of Tab.~\ref{TabRes}.
\label{fig:waves2}}
\end{center}
\end{figure*}

Figure~\ref{fig:waves2} shows two butterfly diagrams for the same model. The toroidal field is shown on the left panel at the same depth as in Fig.~\ref{fig:waves1} and the poloidal field is shown at the surface in the right panel. These two panels demonstrate that the wave-like motion is also observed at the surface of our models, which would result in a Sun-like butterfly diagram. Following \cite{Gilman83}, we also plotted here the axisymmetric toroidal components of the kinetic (\noc{dashed line}) and magnetic (\noc{solid line}) energies, normalized by the total kinetic and magnetic energies, respectively. As \cite{Gilman83} reported, it is possible to identify the imprint of the wave in the kinetic and magnetic energy time series.

A detailed study of dynamo cases is out of the scope of this paper, since we intended to focus \noc{simply on the mechanisms that alter} the kinetic helicity and with it the direction of migration. Aspects related to the magnetic field will be further analysed in future work. For some of these cases, the dynamo wave is not stable, eventually stabilizing in an octupolar solution, which may be related to the use of variable transport properties from our work on Jupiter models.

\section{Analysis and interpretation}
\label{results2}

\subsection{Correlations to describe flow behaviour}
\label{correlation}

According to the idealized scenario described in Sec.~\ref{helicity}, columnar convection often dominates in the inner part of the shell, where the `rising/sinking' component of the flow $u_z$ (along a column and away from the equator) is anti-symmetric about the equator but (since the column rotates in the same direction in both hemispheres) $\omega_z$ is symmetric. Converting between cylindrical and spherical coordinate systems,
\begin{equation}
\begin{split}
	u_z = u_r\cos\theta - u_\theta\sin\theta \Rightarrow 
	u_r = u_s\sin\theta + u_z\cos\theta \textrm{,}
\end{split}
\label{eq:RZcomponents}
\end{equation}
where $s$ represents here the cylindrical radius, we see that the radial component of the velocity $u_r$ is symmetric over the equator, while $\omega_r$ is anti-symmetric. The radial quantities are particularly relevant in strongly stratified models, where convection becomes plume-like thus the preferred 'rising/sinking' direction is along spherical radius. At each radial level, divergence and convergence of the flow is represented by the horizontal divergence of the flow (in spherical coordinates)
\begin{equation}
\begin{split}
	\nabla_h\cdot u_h = 
		\frac{1}{r\sin\theta} \frac{\partial}{\partial\theta} (u_\theta\sin\theta) +
		\frac{1}{r\sin\theta} \frac{\partial u_\phi}{\partial\theta}
	\textrm{.}
\end{split}
\label{eq:divh}
\end{equation}
A similar divergence can be adapted to columnar convection, by assuming a horizontal plane defined by the cylindrical coordinates $s,\phi$ at a fixed height $z$, thus replacing $u_h\!=\!f(\theta,\phi)$ by $u_{hz}\!=\!f(s,\phi)$ and similarly $\nabla_h$ by $\nabla_{hz}$.

\begin{figure*}
\begin{center}
{\centering
	\includegraphics[trim=-0.1cm 0.2cm 0.0cm 0.1cm, height=6.0in]{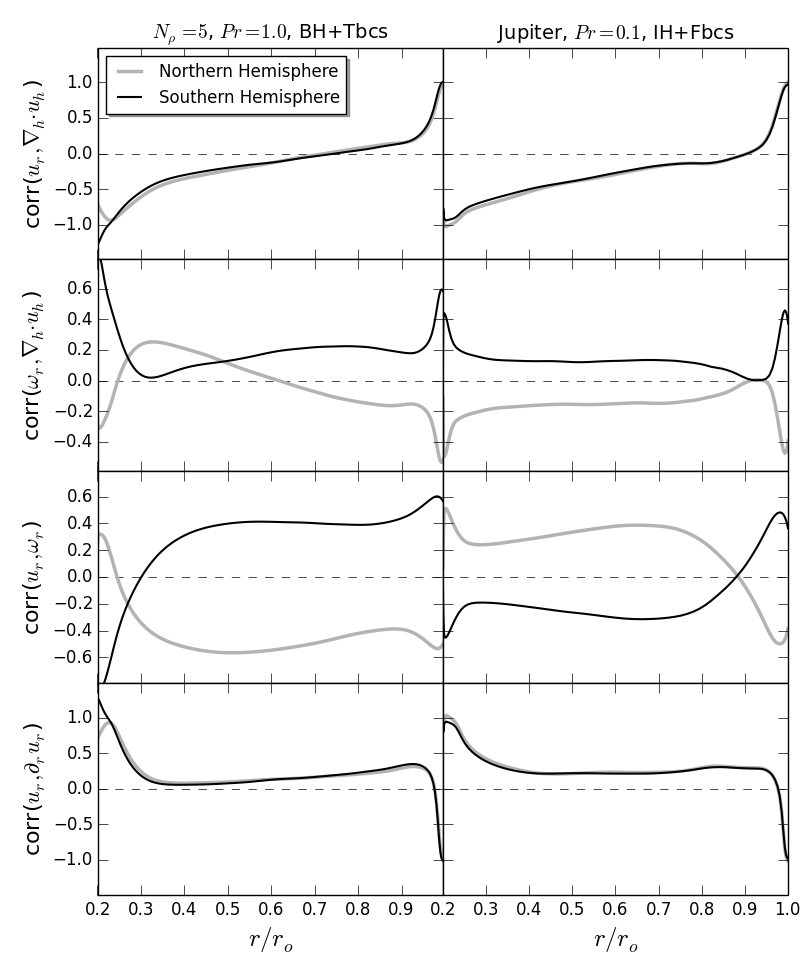}
	}
\caption{\small Correlations between several properties of the flow along the radial direction, namely rising/sinking flow $u_r$, horizontal divergence $\nabla_h\cdot u_h$, radial vorticity $\omega_r$ and acceleration/deceleration $\partial_r u_r$. The left column corresponds to model $4$ in Tab.~\ref{TabRes} and the right column to model $11$. All correlations were calculated at each grid point according to Eq.~\ref{eq:corr}, averaged over the horizontal spherical coordinates $\theta,\phi$ at each radial level $r$ \noc{and over time from $\sim\!100$ snapshots distributed over an interval of $\sim\!20\%$ of a viscous time}. 
\label{fig:corrR}}
\end{center}
\end{figure*}

Figure.~\ref{fig:corrR} shows several correlations between different properties of the rising/sinking flow that dominate convection, only in the radial direction for a first analysis. The model of the panels of the left column represents the helicity set-up described in Sec.~\ref{helicity} and the model of the right column corresponds to the set-up described in Sec.~\ref{helicity2}. Focusing mainly on the right column, since we expect convection to be non-columnar here, in the top panel we see that the rising velocity $u_r$ correlates negatively with the horizontal divergence throughout most of the radius as expected, i.e. as the flow rises it contracts and when it sinks it diverges, as described in Sec.~\ref{helicity2}. The picture naturally inverts in the outer $10\%$ of the layer, where the density gradient dominates, causing a positive correlation between the two quantities. The second-row panel on the right side shows the expected role of the Coriolis force acting on the diverging/converging flow to generate negative/positive $\omega_r$ in the northern hemisphere (\noc{grey} line) and positive/negative in the southern (\noc{black} line). The third panel correlates directly $u_r$ and $\omega_r$, clearly showing that rising/sinking flow in the outer $\sim\!10\%$ of the shell gives negative/positive vorticity, while the picture inverts in the bulk. This panel also translates directly into the preferred sign of kinetic helicity, negative near the surface but positive below $\sim\!90\%$ (Sec.~\ref{helicity2} scenario). Finally, the bottom panel displays the correlation between $u_r$ and the acceleration or deceleration of the fluid, since in the continuity equation it is $\frac{d u_r}{dr}$ that is ultimately linked to horizontal divergence/convergence (if the compressible term is negligible). This shows that the flow is consistently speeding up as it rises as well as slowing down when it sinks in most of the shell, which explains the tendency to contract as it rises.

\begin{figure*}
\begin{center}
{\centering
	\includegraphics[trim=-0.1cm 0.2cm 0.0cm 0.1cm, height=6.0in]{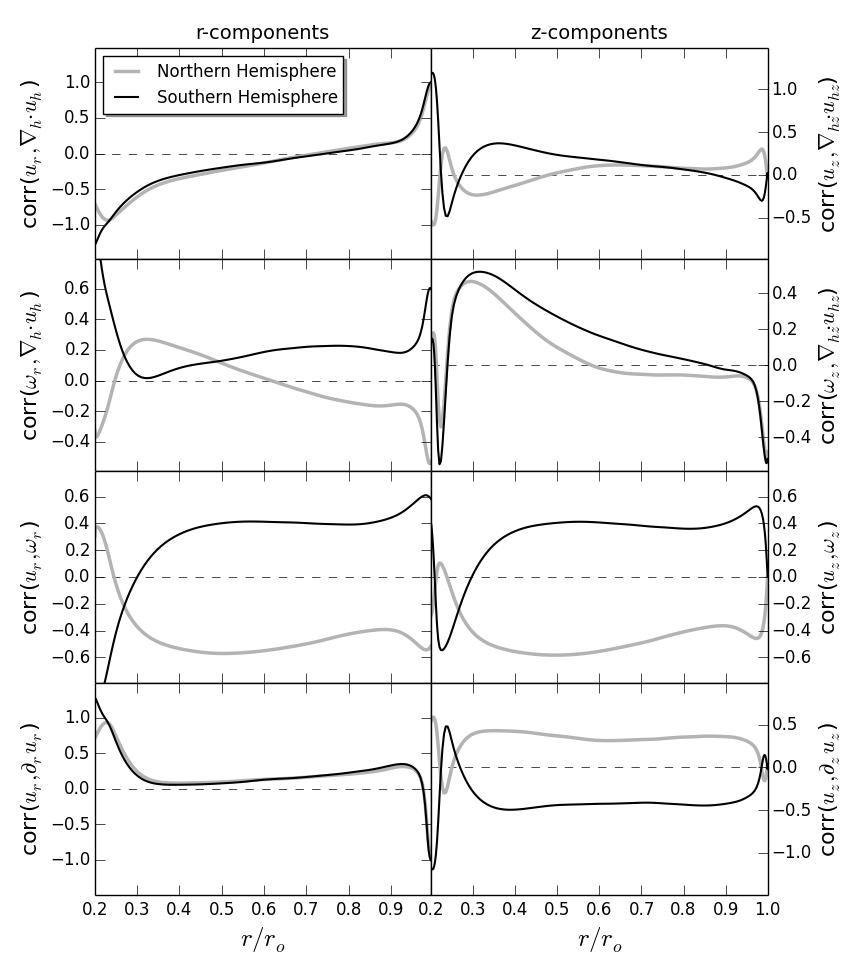}
	}
\caption{\small Correlations between several properties of the flow along the radial direction on the left column (similarly to Fig.~\ref{fig:corrR}) and along the cylindrical $z$ coordinate on the right column. Both columns correspond to model $4$ in Tab.~\ref{TabRes}. \noc{All correlations were calculated at each grid point according to Eq.~\ref{eq:corr}, averaged over the horizontal spherical coordinates $\theta,\phi$ at each radial level $r$ and over time from $102$ snapshots distributed over an interval of $\sim\!20\%$ of a viscous time}.
\label{fig:corrRZ}}
\end{center}
\end{figure*}

The left column of Fig.~\ref{fig:corrR} displays a model which has been shown in the previous Section to be mostly dominated by columnar convection in the bulk, resulting in the helicity pattern described in Sec.~\ref{helicity}. As a result, the radial correlations displayed in the left column of Fig.~\ref{fig:corrR} show some inconsistencies and asymmetries between the two hemispheres, while the right side panels of the same Figure show almost perfect symmetry. This occurs because in columnar convection, such quantities are better expressed in terms of the cylindrical $z$ coordinate. Thus Fig.~\ref{fig:corrR} shows two examples of changes in the flow regime as the rotational constraint is varied, as also seen in the values of convective local Rossby number: $Ro_{\ell\,conv}\!=\!0.276$ and $Ro_{\ell\,conv}\!=\!0.423$ for the models on the left and right columns, respectively.

Figure~\ref{fig:corrRZ} shows the same left column of Fig.~\ref{fig:corrR} and the right column represents equivalent calculations for the same model along $z$, as described in the beginning of this Section. There is some asymmetry in both columns, since there is a small superposition of the two convective regimes, but the right column more clearly illustrates the behaviour of the flow. In the right top panel, when $u_z$ is positive (away from the equator and toward the upper boundary) in the northern hemisphere, the flow will diverge at the top of a column which is seen by tracking the \noc{grey} line. The inner part is an exception though, likely due to either a superposition with the radial component or to the influence of convection inside the TC (not explored here) since the correlations shown are averaged over horizontal spherical coordinates (see caption of Fig.~\ref{fig:corrR}). An asymmetry is also perceptible in the second row, though the cylindrical coordinate system plays the major role in contributing the positive peak near the bottom of the layer ($0.3-0.4\,r_o$) in the radial correlations between $\omega_r$ and $\nabla_h u_h$ in the northern hemisphere (\noc{grey} line). As illustrated above in Fig.~\ref{fig:column2}, the flow diverges mainly at the extremities of a column with negative $\omega_z$, where it encounters the bottom and top boundaries, but it converges along its length in most of the bulk. This explains the positive peak of $\textrm{corr}(\omega_z,\nabla_{hz} u_{hz})$ in the bottom half of the shell that directly affects the sign of $\textrm{corr}(\omega_r,\nabla_h u_h)$ through Eq.~\ref{eq:RZcomponents}, appearing to result in an inconsistent behaviour of the Coriolis force in the radial component. It's the cylindrical calculation on the right panel of the second row that is most relevant for this part of the shell. The top four panels show this persistent asymmetry between the two hemispheres as a direct consequence of the different symmetries between $u_r$/$u_z$ and $\omega_r$/$\omega_z$. This asymmetry is also seen in the bottom four panels, though more camouflaged. In particular the third row once again shows the clear resulting sign of kinetic helicity in the majority of the shell, consistent with the set-up described in Sec.~\ref{helicity}.

\begin{figure}
\begin{center}
{\centering
	\includegraphics[trim=1cm 0.5cm 1cm 0.5cm, height=1.35in]{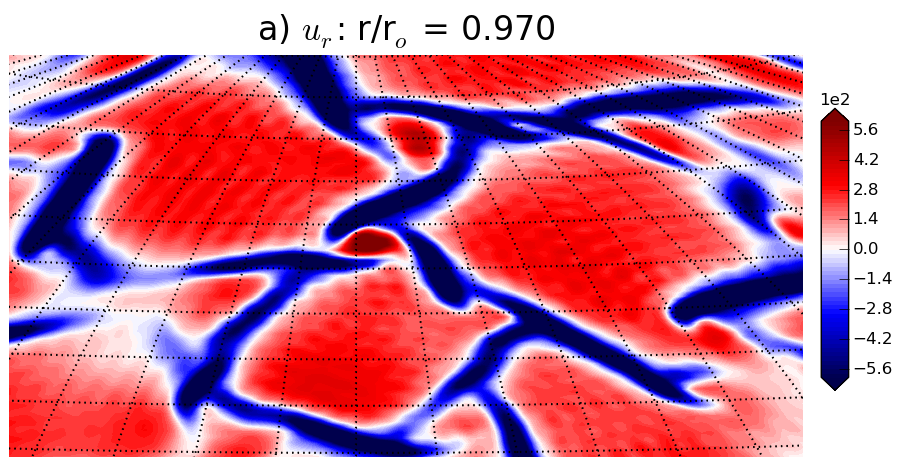}
	\includegraphics[trim=1cm 0.5cm 1cm 0.5cm, height=1.35in]{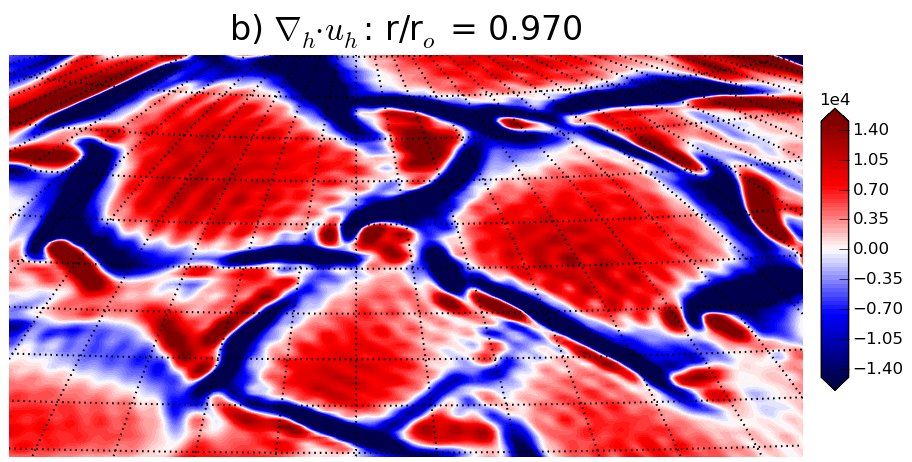}
	\includegraphics[trim=1cm 0.5cm 1cm 0.5cm, height=1.35in]{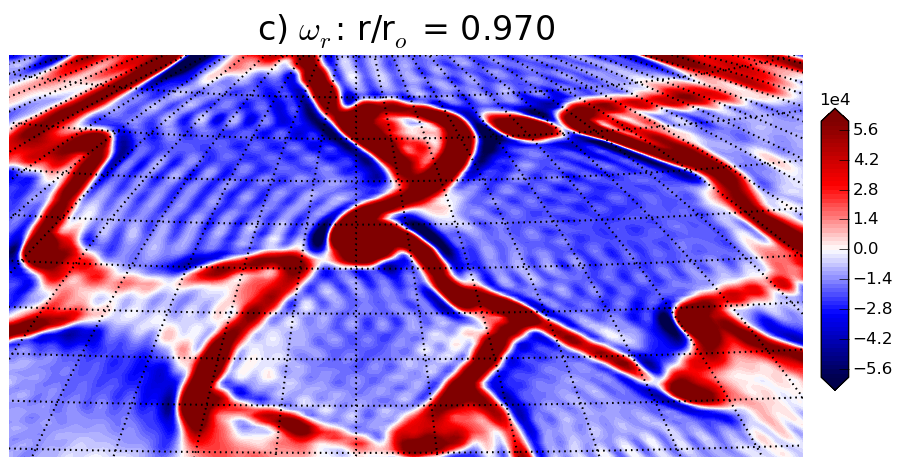}
	}
\caption{\small Properties of the flow in the upper layer of stronger density gradient in the northern hemisphere, in a segment extracted from the right bottom panel of Fig.~\ref{fig:kinhel} (model $11$ of Tab.~\ref{TabRes}). The radial level corresponds to the outer white dashed line in the same Figure ($r\!=\!0.97r_o$). The dotted black meridians and latitude circles are 3 degrees apart.
\label{fig:zoomflow97}}
\end{center}
\end{figure}

\begin{figure}
\begin{center}
{\centering
	\includegraphics[trim=1cm 0.5cm 1cm 0.5cm, height=1.35in]{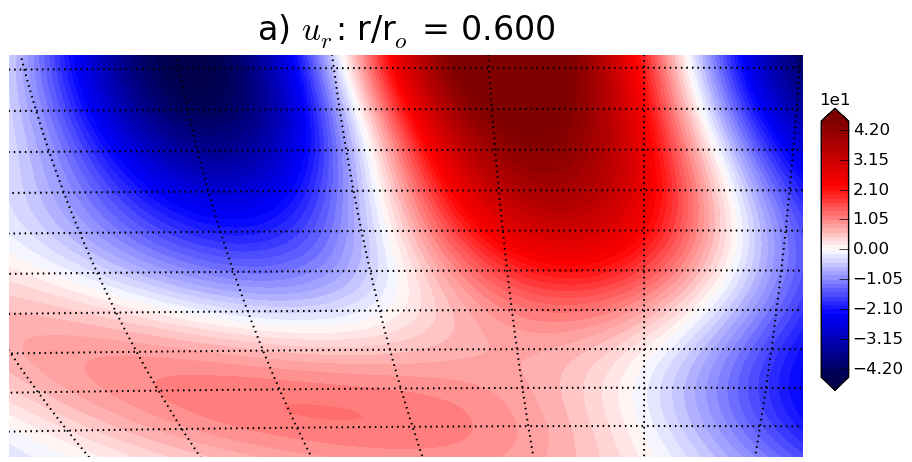}
	\includegraphics[trim=1cm 0.5cm 1cm 0.5cm, height=1.35in]{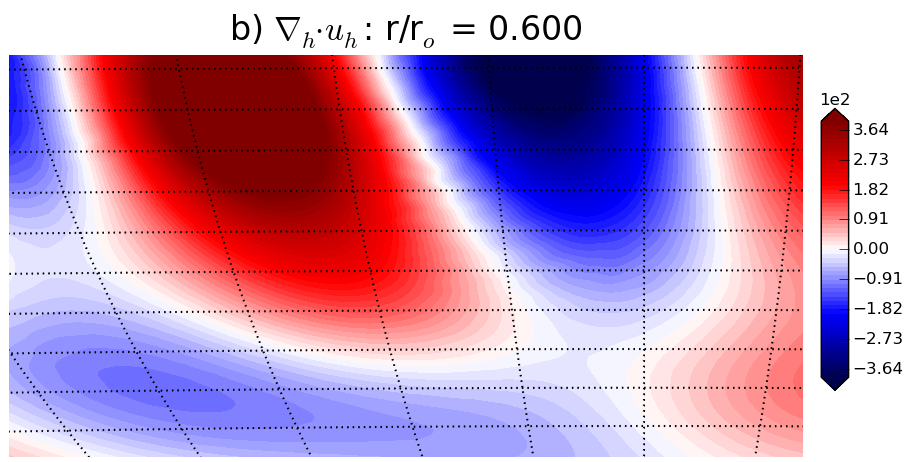}
	\includegraphics[trim=1cm 0.5cm 1cm 0.5cm, height=1.35in]{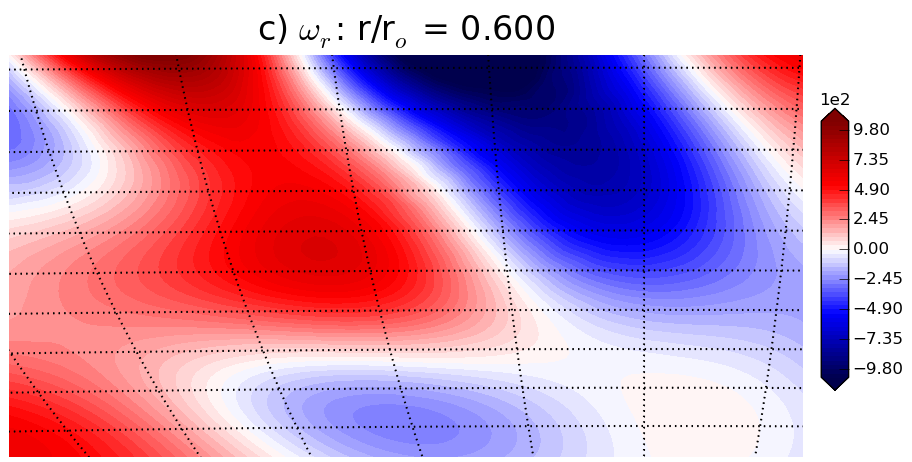}
	}
\caption{\small Properties of the flow in the lower layer of milder density gradient and inverted helicity in the southern hemisphere for model $16$ of Tab.~\ref{TabRes}. The radial level now corresponds to $r\!=\!0.6r_o$.
\label{fig:zoomflow80}}
\end{center}
\end{figure}

Figure~\ref{fig:zoomflow97} shows a segment of the near-surface maps of different properties of the flow to illustrate the typical behaviour of the flows at the outer radii common for most strongly stratified models. In the northern hemisphere, the flow rises (first panel from the top), expands as the density decreases (horizontal divergence of the horizontal velocity $\nabla_h\cdot u_h$, in the third panel) and rotates clockwise (negative $\omega_r$ in the second panel, i.e. anti-cyclones -- see Sec.~\ref{helicity}). The same happens when the flow sinks as it is promptly compressed in the filaments seen in these three panels around the sources represented by the horizontal divergence of the horizontal flow $\nabla_h\cdot u_h$.

Figure~\ref{fig:zoomflow80} illustrates the alternative convection mechanism in the inner part of the shell described in Sec.~\ref{helicity2}, at a depth of $40\%$ below the surface of the model. The different dynamics shown above in Fig.~\ref{fig:corrR} by the inversion of sign in the correlations of the third row, can also be seen here as positive $u_r$ in the top panel of Fig.~\ref{fig:zoomflow80} now correlates with negative horizontal divergence (middle panel) and thus negative $\omega_r$ (cyclones, see Sec.~\ref{helicity2}) due to the action of the Coriolis force. However, these linkages are less clear than in Figure~\ref{fig:zoomflow97}, or the idealised description of Sec.~\ref{helicity2}, partly because of the complexity of the flow field and because a variety of processes contribute at some level to vorticity generation. On average, though, as Fig.~\ref{fig:corrR} shows, upflows in this region are well correlated with horizontally convergent flows.

\subsection{Columnarity}
\label{column}

Contour plots of radial velocity shown in Fig.~\ref{fig:vrax} of the previous Section show the predominance of the columnar convection regime in the bulk of a spherical shell, as suggested in Sec.~\ref{helicity2}. Near the surface both set-ups of Secs.~\ref{helicity} and \ref{helicity2} are identical, except for a slight difference in the scale of the convection, since lower value of $Pr$ is known to produce larger convective flow scales \citep[][]{Jones09a}. At deeper radial levels as discussed above, columnar convection tends to remain the dominant type of convection, while still predominantly plume-like near the surface. \cite{Soderlund12} defined a parameter they called ``Columnarity", meant to determine the degree at which the flow is organized in convection columns. They defined this parameter from the ratio of the integral of $\omega_z$ in the cylindrical $z$-direction and the integral in $z$ of the total RMS vorticity $\omega$ of the flow:
\begin{equation}
	C_{\omega z} =
		\frac{\displaystyle\sum_{s,\phi} \lvert\langle \omega'\cdot\hat{\mathbf{z}} \rangle_z\rvert}
		{\displaystyle\sum_{s,\phi} \langle\lvert \omega' \rvert\rangle_z} \textrm{,}
\label{eq:columnarity}
\end{equation}
where primes mean that vorticity is calculated from the non-axisymmetric part of the velocity field.

\begin{figure}
\begin{center}
{\centering
	\includegraphics[trim=0.5cm 0.5cm 0.5cm 0.0cm, height=2.53in]{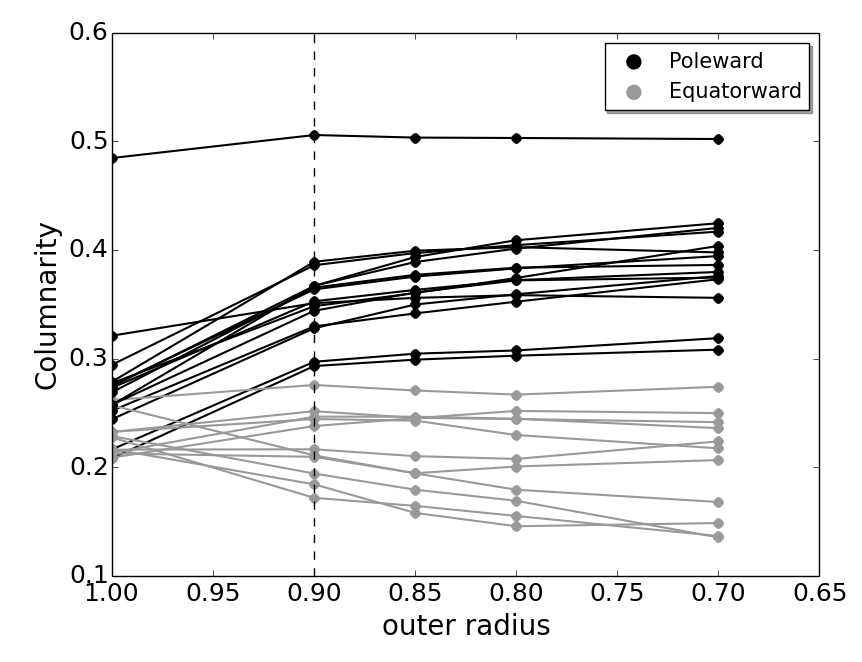}
}
\caption{\small Columnarity \citep{Soderlund12} as a function of the outer radius used as the top boundary for the integration/sum in Eq.~\ref{eq:columnarity}. \noc{ The black lines correspond to models $1$, $2$, $3$, $4$, $5$, $6$, $22$, $23$, $24$, $25$, $26$, $27$, $28$, $29$, $30$ of Tab.~\ref{TabRes},  which have the helicity pattern from Sec.~\ref{helicity} and poleward propagation in the case of magnetic cases. The grey lines correspond to models $7$, $8$, $9$, $10$, $11$, $12$, $16$, $17$, $18$, $19$, $20$, which have the helicity pattern described in Sec.~\ref{helicity2} and equatorward propagation if magnetic. The remaining cases of Tab.~\ref{TabRes} do not have a clear helicity pattern nor a preferred direction of propagation.}
\label{fig:columnarity}}
\end{center}
\end{figure}

This parameter works well for Boussinesq models, where convection onsets and remains attached to the inner boundary, unless the supercriticality is large enough (typically above 50-100 times supercritical). We saw that even though this criterion works fairly well for mildly stratified models (e.g. $N\!\lesssim\!3$), it is not descriptive enough for strongly stratified models. However, the previous Sections suggest that the most likely reason for this is that even though the flow is still columnar in the bulk, it is not in the outer $\sim\!10\%$ of the radius. This will naturally affect the total value of $C_{\omega z}$ since the integration includes this outer part of the shell where convection is not columnar. As a result, the value of $C_{\omega z}$ is nearly the same for models with different helicity patterns. We attempted to calculate $C_{\omega z}$ below deeper radii to remove the major effect of the outer part and thus find a way to still use this parameter to distinguish one-layer helicity models from two-layer. The result is shown in Fig.~\ref{fig:columnarity} for several of our models, obtained from snapshots. By excluding the outer part, it appears indeed possible to separate the two regimes using $C_{\omega z}$, i.e. to separate the poleward-propagating wave models from the equatorward ones. The two different rotational regimes represented by the \noc{grey/black} lines are also characterized by a generally higher value of $Ro_\ell$, as mentioned in previous Section.

\section{Discussion and application to stars}
\label{conclusions}

In many classic theories of stellar and planetary dynamos, the direction of propagation of dynamo ``waves" is determined partly by the differential rotation and partly by the kinetic helicity \citep[e.g.,][]{Parker55,Steenbeck66,Moffatt78}. In broad accordance with this expectation, many prior dynamo simulations that possessed some form of cyclical behavior have exhibited poleward propagation of fields \citep{Gilman83,Ghizaru10,Goudard08,Schrinner11,Simitev12,Gastine12a,Duarte13}, in keeping with the realized profiles of helicity and differential rotation but in contrast to what is seen in the Sun. Notable exceptions in the stellar context include \cite{Kapyla13,Augustson13}, and \cite{Warnecke14}, who attributed equatorward propagation in the simulations primarily to changes in the differential rotation profile. In those simulations, as in prior ones exhibiting poleward propagation, the kinetic helicity of the flows appears to have remained predominantly negative in the northern hemisphere (positive in the southern hemisphere), in accord with theoretical expectations for both columnar and highly stratified convection (as summarized in Sec.~\ref{helicity}). In the planetary context, models with equatorward propagating dynamo waves in a set-up closer to ours were reported by \cite{Jones14} for ``failed" Jupiter-like simulations, carried out at lower values of Prandtl number and always driven by internal heating.

We have examined here whether the kinetic helicity in global-scale convection simulations of stars and planets must necessarily accord with the classic expectation described above (and in Sec.~\ref{helicity}), or whether other self-consistent profiles are possible. We have demonstrated that in some cases equatorward migration of magnetism can arise not from unusual differential rotation profiles, but from the realization of a kinetic helicity profile that is the opposite to that encountered in many prior simulations. We have focused here on the mechanisms by which this helicity ``inversion" is accomplished, while deferring a detailed study of the properties of dynamo solutions to later work.

Our analysis indicates that the ``classic" helicity configuration commonly results when convection is primarily columnar (as often occurs in rapidly rotating cases with small to moderate density contrasts), when no internal heating occurs, and/or when the Prandtl number is unity or greater (as adopted in many prior simulations). But by changing a combination of these factors, we have demonstrated that the helicity in the bulk of the fluid may switch sign, becoming positive (in the northern hemisphere) throughout much of the rotating domain (rather than just in narrow boundary layers). This is due in part to the promotion of a second, deeper layer below the outermost regions with a stronger density gradient. In this deep layer, convective flows are neither columnar nor dominated by expansion/contraction associated with the density gradient. The resulting ``plume-like" flows there tend to have upflows that are associated with converging horizontal flows (rather than divergent ones, as realized in more strongly stratified regions), which in combination with Coriolis forces leads to cyclonic vorticity and positive kinetic helicity. Lowering the fluid Prandtl number promotes the disruption of convective columns due to the first-order effect of inertia on the force balance, but additionally the presence of internal heating in the system and a mild density stratification in the deep interior help to finalize a stable inversion of the helicity pattern.

In summary, to guarantee plume-like convection in the bulk of a spherical domain, the three effects required in our models were strong density stratification (with most of the density gradient confined to the outermost region), lower fluid Prandtl number, \noc{and a combination of} internal heating \noc{and fixed flux boundary conditions} to weaken convection in the bulk. The combination of the last two effects appears to be particularly essential in our cases. \noc{We did not carry out an extensive study of the relative thickness of the two helicity layers in each hemisphere, its dependence on control parameters (including the fluid Prandtl number) or the transitional $Pr$ needed to reverse the sign of helicity. Because $Pr$ is also a function of many other control parameters (including Ekman and Rayleigh numbers and different boundary conditions or heating modes), we consider this to be beyond the scope of this paper.} However, a few preliminary simulations suggest that the size of the outer layer relative to the size of the domain -- in which the helicity remains negative in the northern hemisphere -- decreases significantly with Ekman number. On the other hand, increasing the Rayleigh number (to approach a Rossby number of order unity) tends to act in the opposite direction, decreasing the extent of the region of inverted helicity. Though the simulations considered here are all unstably stratified throughout their interiors, the fraction of energy carried by the convection is small in some cases (primarily because of the low value of $Pr$ adopted); we expect this fraction to grow with Ra, and this may further influence the size of the ``inverted” helicity region. It is not yet clear how these effects would combine to determine the helicity profile at much lower $E$ and much higher $Ra$, but we intend to examine this in future work.

It is not entirely clear whether the ``inverted" kinetic helicity profiles explored here, and the accompanying equatorward migration of magnetic fields, are likely to be realized in stellar or planetary interiors. Some of the factors we have identified as contributing to these profiles are reliably present: for example, $Pr \ll 1$ in many astrophysical plasmas \citep[including the Sun; see, e.g.,][]{Miesch05}. Likewise, the condition that density stratification be comparatively weak throughout part of the interior, so that the expansion/contraction of rising/sinking parcels (and the vortical horizontal flows associated with this) do not utterly dominate the production of kinetic helicity, is satisfied in many stellar and planetary convection zones. Even in the Sun, the density scale height at the base of the convection zone is comparable to the depth below the surface, decreasing rapidly only nearer the photosphere. (Near the photosphere, it seems safe to assume that the expansion/contraction of rising/falling convective cells will dominate over other effects, leading to kinetic helicity that is negative in the northern hemisphere.) On the other hand, we have found that extended internal heating is also important for giving rise to the ``inverted" helicity profiles; in its absence (i.e., in cases heated solely from below), both mixing length theory and our simulations suggest that (in regions where the density stratification is weak) the velocity should not increase significantly with radius outside of a narrow boundary layer. This in turn often leads to the ``classic" helicity profile except in the boundary layer. In essence, the presence of extended internal heating allows a phenomenon that might otherwise be confined to a narrow boundary layer to persist throughout much of the fluid. Because of this, we suspect that our results may be more relevant to objects like brown dwarfs and very low mass stars, in which internal energy generation by fusion or gravitational contraction extends over a large fraction of the interior \citep[e.g.,][]{Chabrier97}, than to stars like the Sun, in which the luminosity that must be carried by convection is roughly constant across the convective envelope. Even in the latter case, however, it is conceivable that non-standard models in which heat transport is dominated by cooling from the top boundary \citep[i.e., ``entropy rain", as studied in][]{Brandenburg15} might lead to kinetic helicity profiles resembling those here. We defer a more detailed exploration of these possibilities, and their consequences for stellar and planetary dynamos, to future work.

{\footnotesize

\onecolumn
\begin{center}
\begin{landscape}
\begin{longtable}{ccccccccccccccccc}

\caption{Summary of mostly time-averaged results, for $E=10^{-4}$. The supercriticality values $Ra/Ra_{cr}$ correspond to the definition applied to non-dimensional formulation \citep[see][for a full description]{Jones09a}. The quantities $C_{\omega z,\,total}$ and $C_{\omega z,\,90\%}$ correspond to snapshots of the values of columnarity determined for the whole shell and below $90\%$ of the total radius, respectively. \\
$^{1}\,Ra_{cr}\!=\!4.648\times 10^6$ \\
$^{2}\,Ra_{cr}\!=\!5.372\times 10^6$ \\
$^{3}\,Ra_{cr}\!=\!2.975\times 10^7$ \\
$^{4}\,Ra_{cr}\!=\!5.412\times 10^6$ \\
$^*$Models where $\tilde{\lambda}$ is a function of radius \citep[see][]{Gastine14}.
\small{
\label{TabRes}}
}\\

\hline
\vspace{-6.5pt}
 & & & \\

Model & $N_{\rho}$ & $n$ & $Ra$ & $Pr$ & $Pm_i$ & $\epsilon$ & $s_{BC,\,i}$ & $s_{BC,\,o}$ & $E_{kin}$ & $E_{mag}$ & $f_{dip}$ & $SD_{dip}$ & $C_{\omega z,\,total}$ & $C_{\omega z,\,90\%}$ & $Ro$ & $Ro_{\ell,\,conv}$ \\
\vspace{-7.5pt}
 & & & \\
\hline
\vspace{-8.5pt}
 & & & \\
\hline
\endfirsthead

\hline
Model & $N_{\rho}$ & $n$ & $Ra$ & $Pr$ & $Pm_i$ & $\epsilon$ & $s_{BC,\,i}$ & $s_{BC,\,o}$ & $E_{kin}$ & $E_{mag}$ & $f_{dip}$ & $SD_{dip}$ & $C_{\omega z,\,total}$ & $C_{\omega z,\,90\%}$ & $Ro$ & $Ro_{\ell,\,conv}$ \\
\vspace{-7.5pt}
 & & & \\
\hline
\vspace{-8.5pt}
 & & & \\
\hline
\endhead

\hline
\multicolumn{15}{r}{{Continued on next page}} \\
\hline
\endfoot
\endlastfoot

\vspace{-8pt}
 & & & \\

1  & 3.0 & 2.0 & ${1.1\!\times\!10^{7}}^{(1)}$ & 1.0 & 2.0 & 0.0 & 0 & 0 & $7.94\!\times\!10^{4}$ & $4.45\times 10^{4}$ & $2.86\!\times\!10^{-3}$ & $4.92\!\times\!10^{-3}$ & $0.485$ & $0.506$ & $0.0078$ & $0.057$ \\ 
2  & 3.0 & 2.0 & ${2.0\!\times\!10^{7}}^{(1)}$ & 1.0 & 2.0 & 0.0 & 0 & 0 & $4.88\!\times\!10^{5}$ & $3.39\!\times\!10^{5}$ & $1.17\!\times\!10^{-2}$ & $6.27\!\times\!10^{-2}$ & $0.321$ & $0.351$ & $0.0186$ & $0.143$ \\ 

\hline

3  & 5.0 & 2.0 & ${4.0\!\times\!10^{7}}^{(2)}$ & 1.0 & 2.0 & 0.0 & 0 & 0 & $1.07\!\times\!10^{6}$ & $9.02\!\times\!10^{5}$ & $3.18\!\times\!10^{-3}$ & $1.31\!\times\!10^{-2}$ & $0.275$ & $0.367$ & $0.0184$ & $0.231$ \\ 
4  & 5.0 & 2.0 & ${5.0\!\times\!10^{7}}^{(2)}$ & 1.0 & 2.0 & 0.0 & 0 & 0 & $1.65\!\times\!10^{6}$ & $1.30\!\times\!10^{6}$ & $2.89\!\times\!10^{-3}$ & $3.17\!\times\!10^{-2}$ & $0.244$ & $0.367$ & $0.0217$ & $0.276$ \\ 

\hline

5$^*$  & 4.9 & 2.2 & ${2.2\!\times\!10^{8}}^{(3)}$ & 1.0 & 1.0 & $0.0$ & 0 & 0 & $4.46\!\times\!10^{6}$ & $7.68\!\times\!10^{5}$ & $3.02\!\times\!10^{-1}$ & $2.61\!\times\!10^{-1}$ & $0.252$ & $0.330$ & $0.0275$ & $0.276$ \\ 

6  & 4.9 & 2.2 & ${4.0\!\times\!10^{8}}^{(3)}$ & 1.0 & 2.0 & $0.0$ & 0 & 0 & $8.34\!\times\!10^{6}$ & $3.16\!\times\!10^{6}$ & $2.91\!\times\!10^{-3}$ & $2.31\!\times\!10^{-2}$ & $0.208$ & $0.293$ & $0.0331$ & $0.400$ \\ 
7  & 4.9 & 2.2 & ${2.6\!\times\!10^{9}}$ & 0.1 & 0.0 & $0.528$ & 1 & 1 & $4.76\!\times\!10^{6}$ & $0.0$ & $-$ & $-$ & $0.227$ & $0.172$ & $0.0322$ & $0.234$ \\ 
8 & 4.9 & 2.2 & ${2.6\!\times\!10^{9}}$ & 0.1 & 4.0 & $0.528$ & 1 & 1 & $3.67\!\times\!10^{6}$ & $6.39\!\times\!10^{5}$ & $1.55\!\times\!10^{-4}$ & $1.90\!\times\!10^{-2}$ & $0.229$ & $0.194$ & $0.0284$ & $0.225$ \\ 
9 & 4.9 & 2.2 & ${3.0\!\times\!10^{9}}$ & 0.1 & 2.0 & $0.528$ & 1 & 1 & $7.91\!\times\!10^{6}$ & $1.25\!\times\!10^{4}$ & $2.22\!\times\!10^{-4}$ & $2.94\!\times\!10^{-2}$ & $0.217$ & $0.184$ & $0.0384$ & $0.308$ \\ 
10 & 4.9 & 2.2 & ${4.0\!\times\!10^{9}}$ & 0.1 & 0.0 & $0.528$ & 1 & 1 & $1.94\!\times\!10^{7}$ & $0.0$ & $-$ & $-$ & $0.213$ & $0.230$ & $0.0584$ & $0.458$ \\ 
11 & 4.9 & 2.2 & ${4.0\!\times\!10^{9}}$ & 0.1 & 2.0 & $0.528$ & 1 & 1 & $8.30\!\times\!10^{6}$ & $2.46\!\times\!10^{6}$ & $5.03\!\times\!10^{-3}$ & $8.38\!\times\!10^{-2}$ & $0.233$ & $0.244$ & $0.0432$ & $0.423$ \\ 
12 & 4.9 & 2.2 & ${5.0\!\times\!10^{9}}$ & 0.1 & 1.0 & $0.528$ & 1 & 1 & $3.44\!\times\!10^{7}$ & $3.26\!\times\!10^{4}$ & $3.39\!\times\!10^{-4}$ & $8.62\!\times\!10^{-3}$ & $0.216$ & $0.217$ & $0.0679$ & $0.582$ \\ 
13$^*$  & 4.9 & 2.2 & ${3.7\!\times\!10^{7}}^{(4)}$ & 0.1 & 4.0 & $0.0$ & 0 & 0 & $4.86\!\times\!10^{6}$ & $5.21\!\times\!10^{6}$ & $8.36\!\times\!10^{-3}$ & $1.77\!\times\!10^{-2}$ & $0.366$ & $0.402$ & $0.0266$ & $0.132$ \\ 
14$^*$  & 4.9 & 2.2 & ${4.0\!\times\!10^{7}}^{(4)}$ & 0.1 & 4.0 & $0.0$ & 0 & 0 & $5.97\!\times\!10^{6}$ & $5.87\!\times\!10^{6}$ & $3.43\!\times\!10^{-3}$ & $4.82\!\times\!10^{-3}$ & $0.358$ & $0.403$ & $0.0305$ & $0.190$ \\ 
15$^*$  & 4.9 & 2.2 & ${6.0\!\times\!10^{7}}^{(4)}$ & 0.1 & 2.0 & $0.0$ & 0 & 0 & $1.60\!\times\!10^{7}$ & $8.96\!\times\!10^{6}$ & $6.50\!\times\!10^{-2}$ & $7.41\!\times\!10^{-2}$ & $0.268$ & $0.312$ & $0.0514$ & $0.422$ \\ 
16$^*$  & 4.9 & 2.2 & ${2.2\!\times\!10^{9}}$ & 0.1 & 20.0 & $0.528$ & 1 & 1 & $2.83\!\times\!10^{6}$ & $5.25\!\times\!10^{3}$ & $2.61\!\times\!10^{-3}$ & $6.75\!\times\!10^{-3}$ & $0.257$ & $0.211$ & $0.0254$ & $0.158$ \\ 
17$^*$  & 4.9 & 2.2 & ${2.6\!\times\!10^{9}}$ & 0.1 & 4.0 & $0.528$ & 1 & 1 & $3.83\!\times\!10^{6}$ & $1.04\!\times\!10^{6}$ & $3.60\!\times\!10^{-1}$ & $2.27\!\times\!10^{-1}$ & $0.232$ & $0.252$ & $0.0311$ & $0.266$ \\ 
18$^*$  & 4.9 & 2.2 & ${2.6\!\times\!10^{9}}$ & 0.1 & 8.0 & $0.528$ & 1 & 1 & $3.69\!\times\!10^{6}$ & $1.16\!\times\!10^{6}$ & $2.33\!\times\!10^{-1}$ & $1.63\!\times\!10^{-1}$ & $0.262$ & $0.276$ & $0.0306$ & $0.267$ \\ 
19$^*$  & 4.9 & 2.2 & ${4.0\!\times\!10^{9}}$ & 0.1 & 2.0 & $0.528$ & 1 & 1 & $9.26\!\times\!10^{6}$ & $2.49\!\times\!10^{6}$ & $4.10\!\times\!10^{-1}$ & $2.81\!\times\!10^{-1}$ & $0.213$ & $0.247$ & $0.0473$ & $0.493$ \\ 
20$^*$  & 4.9 & 2.2 & ${5.0\!\times\!10^{9}}$ & 0.1 & 2.0 & $0.528$ & 1 & 1 & $1.66\!\times\!10^{7}$ & $2.30\!\times\!10^{6}$ & $1.13\!\times\!10^{-1}$ & $1.33\!\times\!10^{-1}$ & $0.209$ & $0.238$ & $0.0557$ & $0.557$ \\ 
21$^*$  & 4.9 & 2.2 & ${4.3\!\times\!10^{7}}$ & 0.1 & 2.0 & $1.0$ & 0 & 0 & $7.27\!\times\!10^{6}$ & $3.33\!\times\!10^{6}$ & $3.78\!\times\!10^{-1}$ & $2.72\!\times\!10^{-1}$ & $0.310$ & $0.338$ & $0.0347$ & $0.198$ \\ 
22  & 4.9 & 2.2 & ${1.4\!\times\!10^{10}}$ & 1.0 & 8.0 & $0.0528$ & 1 & 1 & $4.90\!\times\!10^{5}$ & $5.55\!\times\!10^{5}$ & $1.26\!\times\!10^{-3}$ & $3.88\!\times\!10^{-3}$ & $0.275$ & $0.364$ & $0.0116$ & $0.174$ \\ 
23  & 4.9 & 2.2 & ${1.8\!\times\!10^{10}}$ & 1.0 & 2.0 & $0.0528$ & 1 & 1 & $9.30\!\times\!10^{5}$ & $5.60\!\times\!10^{5}$ & $4.48\!\times\!10^{-4}$ & $1.89\!\times\!10^{-3}$ & $0.257$ & $0.367$ & $0.0150$ & $0.216$ \\ 
24$^*$  & 4.9 & 2.2 & ${1.5\!\times\!10^{10}}$ & 1.0 & 8.0 & $0.0528$ & 1 & 1 & $1.62\!\times\!10^{6}$ & $3.54\!\times\!10^{3}$ & $9.70\!\times\!10^{-2}$ & $6.69\!\times\!10^{-2}$ & $0.273$ & $0.353$ & $0.0181$ & $0.212$ \\ 
25$^*$  & 4.9 & 2.2 & ${1.8\!\times\!10^{10}}$ & 1.0 & 8.0 & $0.0528$ & 1 & 1 & $2.21\!\times\!10^{6}$ & $4.68\!\times\!10^{4}$ & $2.97\!\times\!10^{-2}$ & $7.61\!\times\!10^{-2}$ & $0.269$ & $0.365$ & $0.0198$ & $0.240$ \\
26$^*$  & 4.9 & 2.2 & ${2.0\!\times\!10^{10}}$ & 1.0 & 6.0 & $0.0528$ & 1 & 1 & $2.92\!\times\!10^{6}$ & $3.26\!\times\!10^{4}$ & $1.62\!\times\!10^{-2}$ & $3.84\!\times\!10^{-2}$ & $0.258$ & $0.344$ & $0.0209$ & $0.257$ \\ 
27$^*$  & 4.9 & 2.2 & ${1.7\!\times\!10^{8}}$ & 1.0 & 2.0 & $1.0$ & 0 & 0 & $1.70\!\times\!10^{6}$ & $4.32\!\times\!10^{5}$ & $1.83\!\times\!10^{-1}$ & $1.34\!\times\!10^{-1}$ & $0.294$ & $0.386$ & $0.0187$ & $0.204$ \\ 
28$^*$  & 4.9 & 2.2 & ${1.7\!\times\!10^{8}}$ & 1.0 & 4.0 & $1.0$ & 0 & 0 & $1.57\!\times\!10^{6}$ & $5.79\!\times\!10^{5}$ & $2.02\!\times\!10^{-1}$ & $1.85\!\times\!10^{-1}$ & $0.277$ & $0.348$ & $0.0180$ & $0.220$ \\ 

29  & 4.9 & 2.2 & ${2.0\!\times\!10^{8}}^{(3)}$ & 1.0 & 2.0 & $0.0$ & 0 & 0 & $2.04\!\times\!10^{6}$ & $1.58\!\times\!10^{6}$ & $1.73\!\times\!10^{-2}$ & $6.16\!\times\!10^{-2}$ & $0.279$ & $0.389$ & $0.0182$ & $0.229$ \\ 
30  & 4.9 & 2.2 & ${3.0\!\times\!10^{8}}^{(3)}$ & 1.0 & 2.0 & $0.0$ & 0 & 0 & $5.01\!\times\!10^{6}$ & $2.18\!\times\!10^{6}$ & $1.31\!\times\!10^{-2}$ & $8.52\!\times\!10^{-2}$ & $0.216$ & $0.297$ & $0.0264$ & $0.335$ \\ 

\hline

31$^*$  & 0.0 & $-$ & ${2.0\!\times\!10^{7}}$ & 0.1 & 2.0 & $0.242$ & 1 & 1 & $3.67\!\times\!10^{6}$ & $9.18\!\times\!10^{5}$ & $9.69\!\times\!10^{-4}$ & $8.27\!\times\!10^{-3}$ & $0.191$ & $0.196$ & $0.0950$ & $0.354$ \\ 
32$^*$  & 0.0 & $-$ & ${3.0\!\times\!10^{7}}$ & 0.1 & 4.0 & $0.242$ & 1 & 1 & $5.34\!\times\!10^{6}$ & $1.25\!\times\!10^{6}$ & $3.02\!\times\!10^{-3}$ & $4.04\!\times\!10^{-2}$ & $0.152$ & $0.153$ & $0.0115$ & $0.459$ \\ 

\hline

\end{longtable}
\end{landscape}
\end{center}
\twocolumn

}

\section*{Acknowledgements}

This work was supported in part by the European Research Council under ERC grant agreement no 337705 (CHASM) and by a Consolidated Grant from the UK STFC (ST/J001627/1). Most of the computations for this paper were carried out in the GWDG computer facilities in G\"ottingen and some were performed on the University of Exeter supercomputer, a DiRAC Facility jointly funded by STFC, the Large Facilities Capital Fund
of BIS and the University of Exeter.

The authors would like to thank Rakesh Yadav for useful discussions and help with the display of Figures and the reviewer for the input which contributed in improving our paper.


\bibliographystyle{mnras}
\bibliography{biblio}


\end{document}